\documentclass[journal]{IEEEtran}
\usepackage{graphicx}
\usepackage[resetlabels,labeled]{multibib}
\usepackage{physics}
\usepackage{amssymb}
\usepackage[font=scriptsize]{caption}
\usepackage{amsmath}
\usepackage{mathtools}
\usepackage{lipsum,adjustbox}
\usepackage{color}
\usepackage{tabularx,booktabs}
\usepackage{subcaption}
\usepackage[T1]{fontenc}
\usepackage[utf8x]{inputenc}
\usepackage{authblk}
\usepackage{multirow}
\usepackage{cite}
\usepackage{hyperref}
\usepackage{lettrine}
\usepackage{longtable}
\usepackage{array}
\newcolumntype{A}{>{\centering\arraybackslash}m{0.2cm}}
\newcolumntype{B}{>{\centering\arraybackslash}m{3.5cm}}
\newcolumntype{C}{>{\centering\arraybackslash}m{3.5cm}}
\newcolumntype{D}{>{\centering\arraybackslash}m{0.6cm}}
\newcolumntype{E}{>{\centering\arraybackslash}m{1cm}}
\newcolumntype{G}{>{\centering\arraybackslash}m{1.2cm}}
 
 \hypersetup{
    colorlinks=false,
    linkcolor=blue,
    filecolor=magenta,
    urlcolor=cyan,
}

\begin{document}

\ifCLASSINFOpdf

\else
 
\fi

\title{Soft Computing Techniques for Dependable Cyber-Physical Systems}
%
%
%

\author[1]{Muhammad Atif}
\author[1,3]{Siddique Latif}
\author[1]{Rizwan Ahmad}
\author[1,2]{Adnan Khalid Kiani}
\author[3]{Junaid Qadir}
\author[4]{Adeel Baig}
\author[5]{Hisao Ishibuchi}
\author[3]{Waseem Abbas}

\affil[1]{National University of Sciences and Technology (NUST), Pakistan}
\affil[2]{University of Hull, United Kingdom}
\affil[3]{Information Technology University (ITU)-Punjab, Pakistan}
\affil[4]{Al Yamamah University,  Kingdom of Saudi Arabia} 
\affil[5]{Southern University of Science and Technology (SUSTech), China}

\maketitle

\begin{abstract}

Cyber-Physical Systems (CPS) allow us to manipulate objects in the physical world by providing a communication bridge between computation and actuation elements. In the current scheme of things, this sought-after control is marred by limitations inherent in the underlying communication network(s) as well as by the uncertainty found in the physical world. These limitations hamper fine-grained control of elements that may be separated by large-scale distances. In this regard, soft computing is an emerging paradigm that can help to overcome the vulnerabilities, and unreliability of CPS by using techniques including fuzzy systems, neural network, evolutionary computation, probabilistic reasoning and rough sets. In this paper, we present a comprehensive contemporary review of soft computing techniques for CPS dependability modeling, analysis, and improvement. This paper provides an overview of CPS applications, explores the foundations of dependability engineering, and highlights the potential role of soft computing techniques for CPS dependability with various case studies, while identifying common pitfalls and future directions. In addition, this paper provides a comprehensive survey on the use of various soft computing techniques for making CPS dependable. 

\end{abstract}


\section{Introduction}
  
\lettrine[findent=1.5pt]{\textbf{T}}HE internet has transformed human life in all sorts of beneficial ways. It has become an indispensable tool for all kinds of operations in the fields of business, manufacturing, trade, education, and services. Despite the ubiquity of advanced high-speed data networks, there still is a gap between the cyber world, in which information is processed or exchanged, and the physical world we inhabit \cite{sanislav2012cyber}. This motivates a vision of Cyber-Physical Systems (CPS) that will integrate computational resources into the physical world \cite{liu2017review} to allow for better control over processes that generate and use information. A CPS can be envisioned as the orchestration of computers and physical systems in which embedded computers monitor and control physical processes, typically through feedback loops, and physical process and computations interact with each other closely \cite{lee2015past}.

CPS emerged as an engineering discipline around 2006 although its intellectual roots date back considerably \cite{CPSWebPage}. The terms ``cyber space'', ``cyber-physical systems'' share a common root with the term ``cybernetics'' that was coined by the influential American mathematician Norbert Weiner in the 1940s as the name of a new field that he founded which focused on the conjunction of physical processes, computation, and communication using ideas from control systems theory. As discussed in \cite{lee2015past}, CPS is now an important independent field of engineering that demands its own techniques, theory, methods, and models. The ubiquitous presence of embedded systems and high-speed data networks and the potential benefits of CPS has led some leading thinkers to anticipate that CPS revolution of the 21st century will likely overshadow the IT revolution of $20^{th}$ century \cite{lee2008cyber}. 

The decreasing cost of complex embedded electronics is  ensuring that embedded technology is finding its way into all kinds of everyday products helping realizing a vision of CPS with virtually endless benefits \cite{oks2017application} \cite{loukas2013review}. CPS are already widely being used in utility networks, transportation systems, entertainment business and in a number of industries including healthcare, manufacturing, and services \cite{gunes2014survey}. More generally, one can envision CPS as a very broad field that encompasses a number of modern trends such as Internet of Things (IoT), Machine-to-Machine (M2M), sensor networks, and fog computing. Some prominent CPS applications include the following (a more detailed description follows in the next section): 

\begin{enumerate}

\item the generation of electrical power can be managed better through ``smart grids''; 
\item factories can be operated much more efficiently allowing us to  cut down on greenhouse gas emissions; 
\item autonomous vehicles, aware of other vehicles and obstacles in their vicinity, will allow us to manage urban problems like traffic congestion and to minimize pollution; 
\item self-aware integrated healthcare systems will allow us to provide universal healthcare; and
\item the security of individuals can be improved through intelligent surveillance and monitoring to reduce urban crime and reduce terrorism thread. 

\end{enumerate}


The socioeconomic benefits of CPS technology have been long recognized (dating decades before even the coinage of term CPS) \cite{glushkov1966introduction}. But the true benefits envisioned with CPS have yet to be unleashed \cite{CPSWebPage} due to user apprehension about limitations such as the lack of reliability, predictability, and real-time control of today's computing and networking technologies, which impedes the broad adoption of CPS applications especially for mission-critical applications (such as automotive safety, traffic control, and healthcare). For mission-critical applications, \textit{dependability} and \textit{reliability} assumes paramount importance since CPS must be robust enough to withstand unexpected conditions in communication networks and capable of adapting to subsystem failures \cite{lee2008cyber}. In general, system dependability is often an uncompromisable fundamental requirement of most CPS applications due to the potential of great loss (financial loss or even loss of life). Typically the individual underlying component (hardware) of a CPS is highly reliable, but the overall interconnected network of these components is still vulnerable since it may suffer from deficiencies such as the lack of `temporal semantics', and an inadequate concurrency model. In fact, a failure or an attack on a single component could initiate the cascading failure phenomenon with detrimental consequences for the overall system.


CPS operations are marked by the faster operational time scales, dynamic environments, heterogeneous components, and a large number of mixed initiative interactions \cite{koutsoukos2017sure}. All these factors introduce a certain degree of imprecision and uncertainty in the information required to undertake the necessary computations. Hence, a computational framework that can deal with
all these factors is needed. In this regards, soft computing techniques have emerged as an enabler to make CPS more robust and adaptable. 
Soft computing techniques were invented to overcome the limitations of traditional (`hard') computing techniques that rely on deterministic analytic techniques that aim to exactly solve problems while assuming full knowledge of the parameters involved \cite{chaturvedi2008soft}. Unfortunately, such assumptions are not met in practical real-life systems in which 
imprecision and unavailability of exact prior knowledge is the norm rather than an exception. Soft computing, in strict contrast to hard computing, can work with imprecision, uncertainty, and incomplete information to achieve approximate ``good enough'' solutions to computationally hard problems at lower costs \cite{chakraborty2017application} \cite{zadeh1994soft}. For example, soft computing can use computational intelligence techniques to heuristically solve intractable Non-deterministic Polynomial-time (NP-)complete problems \cite{yu2010introduction} to produce approximate ``good enough'' solutions. A comparison of hard and soft computing is presented in Table \ref{Table II}. 

\begin{table}[!t]
\caption{Hard vs Soft Computing (adapted from \cite{chakraborty2017application})}
\begin{tabular}{ | m{2.2cm} |m{2.4cm} |m{2.7cm}| } 
\hline  
\textbf{Attribute}&\textbf{Hard Computing}&\textbf{Soft Computing}\\ 
 \hline  
Accuracy vs. \newline Robustness&Accuracy Mandatory&Robustness has priority\\
\hline 
Logic&Binary logic&Multi-valued logic\\
\hline 
Input Data&Exact data required&Can work around imprecise data\\
\hline 
Computation Mode&Mostly Sequential&Supports Parallelism\\
\hline 
Precision of results&Precise answers&Approximate answers\\
\hline 
Determinism&Deterministic&Non-deterministic\\
\hline 
\end{tabular} 

\label{Table II}
\end{table}

Various studies, books, and review articles on the scope and applications of CPS are available in existing literature \cite{baheti2011cyber,lee2008cyber,shi2011survey,rajkumar2010cyber}, due to the enormous industrial and scientific research in CPS. 
Similarly the use of soft computing techniques for modeling, analysis and optimization of CPS has been heavily researched in the literature \cite{khan2014computational,khaitan2015design,mitchell2014survey,wan2011advances}. \textit{However, despite the vast literature, a comprehensive survey on the role of soft computing techniques in dependable CPS is missing in the literature.} This is highlighted in Table \ref{survey_summary}, where we compare our survey paper with existing resources in the same space. 

To summarize, the main highlights of our paper are as follows: (1) this paper provides an overview of CPS and their applications in real life; (2) concepts related to the reliability of CPS are introduced in detail; (3) a detailed taxonomy of soft computing techniques is presented; (4) applications of soft computing techniques for modeling, analyzing, and improving the dependability of CPS are discussed; (5) insights are shared on the suitability of various soft computing techniques for various CPS dependability modeling, analysis, and optimization tasks, and finally (6) open issues and directions for future works are identified. 
 
\begin{table}[!ht]
\centering
\caption{List of Abbreviations}
\begin{tabular}{|m{1.1cm} | m{3.4cm} |}
\hline 
 
ACO&Ant Colony Optimization\\
\hline
ANN&Artificial Neural Network \\
\hline
BN&Bayesian Network\\
\hline
CPS&Cyber-physical system\\
\hline
CS&Cuckoo Search\\
\hline
EC&evolutionary computation\\
\hline
FL&Fuzzy Logic\\
\hline
FS&Fuzzy Set\\
\hline
FT&Fault Tree\\
\hline
FTA&Fault Tree Analysis\\
\hline
GA&Genetic Algorithm\\
\hline
IDS&Intrusion Detection System\\
\hline
MLN&Markov Logic Network\\
\hline
MRF&Markov Random Field\\
\hline
NCS&Networked Control System\\
\hline
PN&Petri Net\\
\hline
PR&Probabilistic Reasoning\\
\hline
PSO&Particle Swarm Optimization\\
\hline
RAP&Resource Allocation Problem\\
\hline
RBD&Reliability Block Diagram\\
\hline
RS&Rough Set\\
\hline
RST&Rough Set Theory\\
\hline
SA&Simulated Annealing \\
\hline
SPN&Stochastic Petri Net\\
\hline
SVM&Support Vector Machine\\
\hline
TS&Tabu Search\\
\hline

\end{tabular} 
\label{Table 1}
\end{table}
  
The rest of the paper is organized as follows. In section II, we present application domains of CPS and motivate dependability in CPS by highlighting various attacks in these domains. In section III, a detailed survey of existing soft computing techniques being used to improve or assess the dependability of CPS is presented. Section IV discusses limitations of current research and conclusions from currently available work on the use of soft computing in modeling or improving CPS dependability. Section V presents a few open issues and directions for future work. A list of abbreviations used frequently in the paper is also included (Table \ref{Table 1}).

\begin{table*}[ht]
\centering
\scriptsize
\caption{ Comparison of our survey with existing survey and review papers. (Legend: $\checkmark$
means covered; $\cross$ means not covered;
$\approx$ means partially covered)}
\begin{tabular}{ |m{2.3cm} | m{0.8cm} |m{1.6cm} |m{1.5cm} |m{1.5cm} |m{1.8cm}| m{1.5cm} |}
\hline
\textbf{Surveys/books (Author)}&
\textbf{Year}
& \textbf{Theoretical Foundations (Y/N)}
& \textbf{Applications (Y/N)}  
& \textbf{Dependability Discussed (Y/N)}
& \textbf{Soft Computing Discussed (Y/N)}
& \textbf{Open Issues or Challenges (Y/N)}
\\ \hline

Wan et al. \cite{wan2011advances}&2011&$\checkmark$&$\checkmark$&$\checkmark$&$\cross$&$\checkmark$\\
\hline
Shi et al. \cite{shi2011survey}&2011&$\approx$&$\checkmark$&$\approx$&$\cross$&$\checkmark$\\
\hline
Gunes  et al. \cite{gunes2014survey}&2014&$\checkmark$&$\checkmark$&$\checkmark$&$\cross$&$\checkmark$\\

\hline
Khan et al. \cite{khan2014computational}&2014&$\checkmark$&$\checkmark$&$\checkmark$&$\checkmark$&$\cross$\\
\hline
Mitchell et al. \cite{mitchell2014survey}&2014&$\checkmark$&$\cross$&$\approx$&\checkmark&\checkmark\\
\hline
Khaitan et al. \cite{khaitan2015design}&2015&$\checkmark$&$\checkmark$&$\checkmark$&$\cross$&$\checkmark$\\
\hline
Edward A. Lee \cite{lee2015past}&2015&$\checkmark$&$\cross$&$\approx$&$\cross$&$\checkmark$\\
\hline
Humayed \cite{humayed2017cyber}&2017&$\checkmark$&$\checkmark$&$\cross$&$\cross$&$\checkmark$\\
\hline
Our Survey&2017&$\checkmark$&$\checkmark$&$\checkmark$&$\checkmark$&$\checkmark$\\
\hline
\end{tabular} 
\label{survey_summary}
\end{table*}

\section{Application Domains and Dependability in CPS}

\begin{figure}[!ht]
\centering
\includegraphics[width=2in]{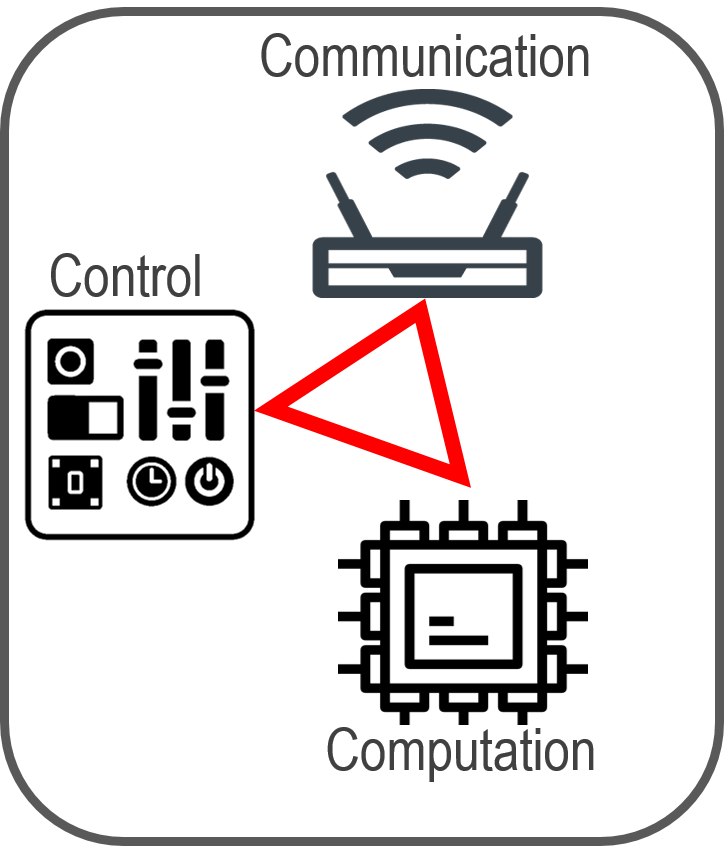} 
\caption{The building blocks of communication, control, and computation subsystems serve as the elements that combine to form cyber-physical systems (CPS).}
\label{figure1}
\end{figure}
\begin{figure*}
\centering
\includegraphics[width=7in]{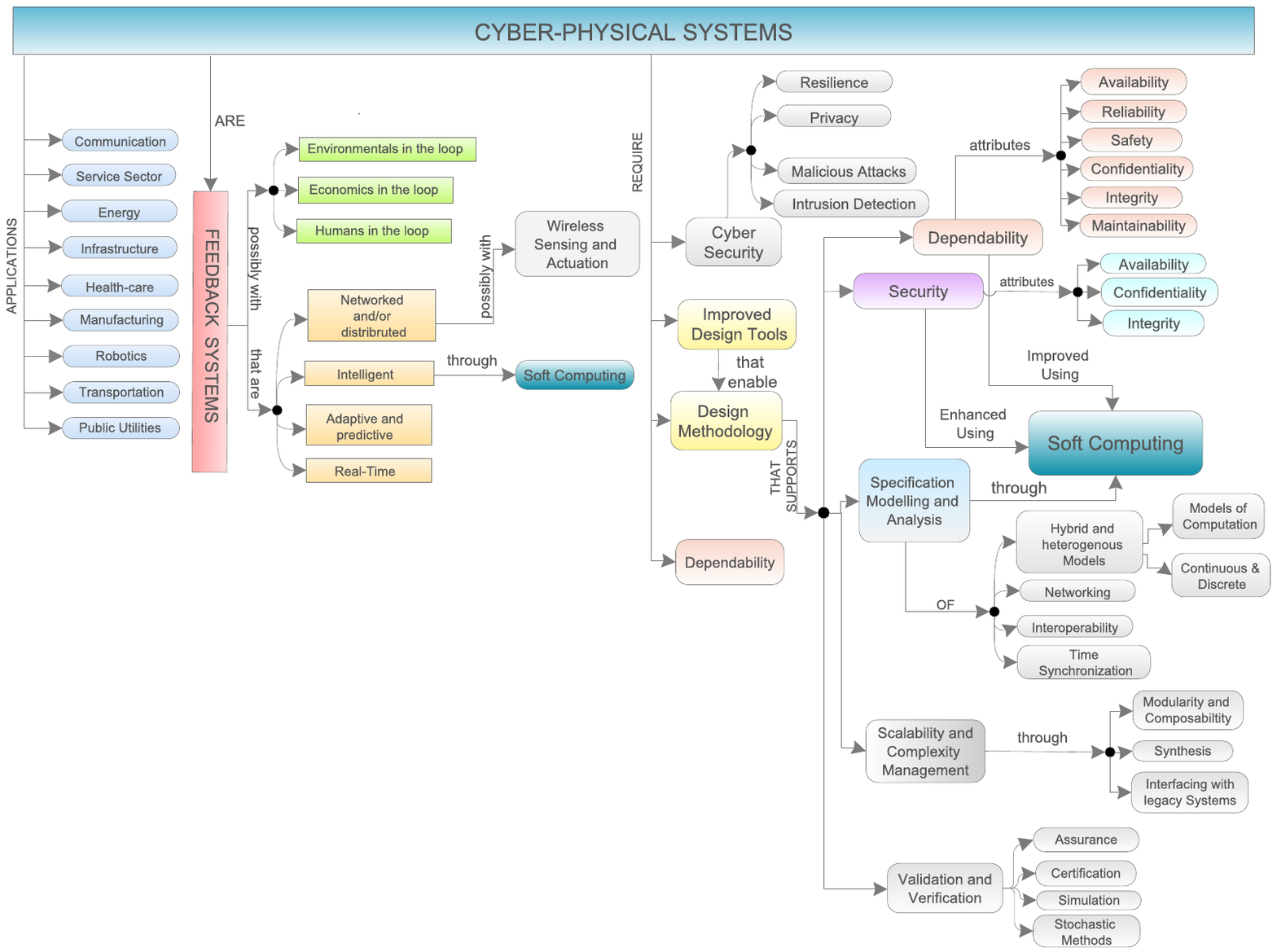} 
\caption{A Concept Map of Cyber-physical Systems (adapted from \cite{CPSWebPage})}
\label{figure2}
\end{figure*}

CPS integrate physical processes with computation and networking. Figure \ref{figure1} shows a typical CPS with the integration of Control, Communication, and Computation. They are sometimes referred to as a Networked Control System (NCS), Distributed Control System (DCS), Sensor Actuator Network (SAN), or Wireless Industrial Sensor Network (WISN) \cite{mitchell2013effect}. It is possible to conceptually model a CPS as a temporally-integrated distributed control system \cite{lee2015past}. CPS allows integration of multiple technologies which have applications spread over several engineering disciplines as highlighted in Figure \ref{figure2}.

\subsection{Application Domains of CPS}

CPS can find uses in almost all fields of modern life. Applications of CPS include transport systems, assisted healthcare, water networks, autonomous vehicles, smart grids for utility networks, and telecommunication, to name a few. In these applications, CPS create an ecosystem where multiple embedded systems can work together towards achieving a common goal. 
This interconnectivity makes CPS more vulnerable to cyber-attacks, cyber-physical attacks, and failure incidents with calamitous consequences in these domains. Below we discuss some common use cases of CPS and different disruptive scenarios in these systems to highlight the need for their dependable and resilient design. 

\subsubsection{Electrical Power Grid (Smart Grids)}

The power grid is a complex and geographically distributed collection of entities that generate, regulate, and utilize power. A combined system of power generation, large-scale distribution and automated power management in the consumer premises, form a CPS. These smart grids provide better fault tolerance, better security, and economic advantages through fine-grained control over the entire system. Smart grids can perform real-time distributed sensing, measurement, and analysis of the production and distribution of electrical power \cite{baheti2011cyber}. Advantages of these technologies include minimization of power outages and reducing greenhouse gas emissions. Despite these benefits, smart grids are vulnerable to cyber and cyber-physical attacks that can cause damages to the whole system at macro scale in terms of the power outage. One such attack happened in Ukraine  in December 2015 \cite{Smart}, in which a targeted cyber-attack against
grid operators resulted in a massive blackout. This left about 225,000 customers without electricity. Such attacks impact the business of smart grid companies and the reliability of their systems.  Hence, there is a need for such platforms that not only secure the individual system components but also strengthen the overall CPS.

\subsubsection{Water Networks}

Water networks are critical infrastructures that have national importance and the quality of normal life directly depends on them. Water networks are very complex consisting of various sensing devices and their complexity is rapidly increasing to meet the rising demands of big cities and industries. Water networks are highly  vulnerable to a variety of attacks. Any cyber or cyber-physical attack can have severe health and economic impacts \cite{ostfeld2008battle}. For instance, in 2000, in Maroochy Shire, Australia, a disgruntled employee
launched a series of attacks on the SCADA system controlling the sewage treatment plant, which resulted in the spillage of 800,000 liters of raw sewage into public
and residential areas causing heavy damages \cite{slay2007lessons}. These instances further illustrate the need to have a secure and a dependable framework to perform CPS operations.

\subsubsection{Industrial Automation} 
CPS can provide a broad control over complex and large industrial facilities through a heterogeneous network architecture of sensors, actuators, and processors \cite{wang2008cyber}. CPS in the industrial chain will result in unprecedented profits for industry and flexibility for consumers \cite{lee2013predictive}. This convergence of automation in the industry with computing and real-time networking is being hailed as the fourth industrial revolution. This has the potential to optimize the entire cycle of production from the supply chain, manufacturing, inventory management, storage, and trade. The ``industrie 4.0'' initiative \cite{lee2015cyber} was taken by the German government to bridge the gap between apparently disparate elements in the supply and production chain. Standards and protocols for communication between often heterogeneous elements in the industrial process are being developed. Introduction of intelligent systems in the industrial automation will make the industry more adaptive to customer requirements. 
On the other hand, these CPS in industrial automation are highly vulnerable. In 2013, foreign hackers, penetrated the control system of a dam located in Rye Brook, New York as a part of larger cyber campaign \cite{Indus1}. Another cyber attack occurred in a German Steel Mill, in 2014, which resulted in monolithic physical damage \cite{millatt}. The adversary gained access to the plant network by using spear phishing email and caused failure of multiple components and critical process of the system.

\subsubsection{Intelligent Transportation Systems (ITS)}
Transport systems are integrating intelligent vehicles with intelligent infrastructure. Context-aware vehicular CPS with cloud support will provide more convenience and safety for drivers, passengers, and pedestrians \cite{wan2014context}. Such systems will minimize urban traffic and parking problems. In a managed transportation system, vehicles can travel together in fleets and the road infrastructure can be used optimally. Smart transportation will assist in times of disaster for emergency evacuation of urban population \cite{rajkumar2010cyber}. Whereas the infrastructure and vehicles required for truly smart transportation systems are in their infancy, the aviation industry is far more mature in terms of technology and communication networks. 
A failure in ITS can cause diverse environmental impacts, wastage of time and make public insecure. Such failures can come from a number of security flaws in the system by designers or due to individual components in ITS. Recently, Ghena et al. \cite{ghena2014green} analyzed the security aspect of a real-world ITS, located in Michigan, to discover different security flaws. They were able to find three major weaknesses in this system. These were lack of encryption, lack of secure authentication, and the existence of vulnerabilities in the software. The authors leveraged these weaknesses and created an attack on the system by showing authorities that an adversary can gain control over traffic infrastructure to gain an unfair advantage by degrading safety and creating disruption.

\subsubsection{Healthcare}
In recent years, CPS are gaining considerable interest for their promising applications in healthcare. Such systems can integrate health monitoring devices such as sensors, actuators, and cameras with cyber components and intelligence. Recently various CPS architectures have been proposed to enhance the healthcare facilities \cite{lee2012challenges}.  In \cite{wang2011secured} a CPS-based secured architecture is presented for healthcare applications that uses the WSN-cloud framework. Similarly, a health-CPS model that consists of the combination of cloud and big data analytics is proposed in \cite{zhang2015health}. The advances in IT and AI will enable CPS-based healthcare systems to provide universal healthcare. As the healthcare CPS offers health services based on the patients' health records or history to improve treatment and patient care \cite{fi9040093}. This personal information in healthcare systems is vulnerable to criminals and cyber threats. An example of such attacks is global ransomware attack that hit the healthcare systems in the United Kingdom, Ukraine, Spain, France and U.S hospitals \cite{Healthcare}.


\subsection{Dependability in CPS}

All of the instances discussed in the previously further emphasize the need to make CPS operations resilient and dependable. Because the applications and services provided by a CPS must be guaranteed and dependable in different environmental contexts (i.e.,  local as well as global). In this section, first, we discuss the notion of dependability in a more general context, thereby discussing it particularly for CPS.

Dependability is a system property that encompasses attributes like ``reliability, availability, survivability, safety, maintainability and security''  \cite{avizienis2001fundamental}. It essentially borrows important concepts from various technologies and merges them into one term \cite{al2009comparative}. International Standards Organization (ISO) defines dependability as \textit{``the collective term used to describe the availability performance and its influencing factors: reliability, performance, maintainability performance and maintenance support performance''} \cite{avizienis2004basic}. The  International Electrotechnical Commission (IEC) defines dependability in terms of percentage of availability \cite{avizienis2004basic}. In computing, dependability is a property of a computing system that allows the user to place reliance on the service it delivers \cite{laprie1995dependable}. An alternate definition for dependability as laid out by the leading researchers in the field is \textit{``the ability to avoid service failures that are more frequent and more severe than is acceptable''}\cite{avizienis2004basic}. The term dependability carries different meanings in different scenarios. The complementary attributes of dependability are highlighted in Figure \ref{Dep_Sec}). which include: 
 \begin{figure}
\centering
\includegraphics[width=3.5in]{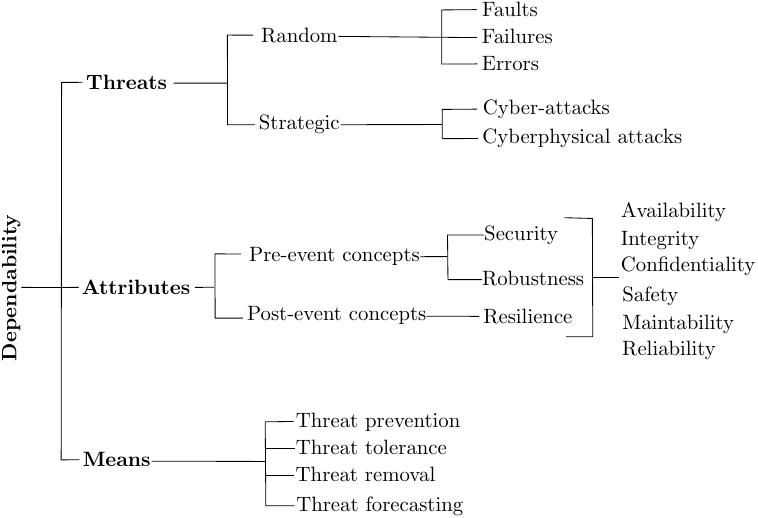} 
\caption{Dependability and Security Attributes.}
\label{Dep_Sec}
\end{figure}

\begin{itemize}
    \item \textit{availability}: readiness for correct service;
    \item \textit{reliability}: continuity of correct service; 
    \item \textit{safety}: absence of catastrophic consequences on the user(s) and the environment;
    \item \textit{confidentiality}: absence of unauthorized disclosure of information;
    \item \textit{integrity}: absence of improper system state alterations; 
    \item \textit{maintainability}: ability to undergo repairs and modifications.
\end{itemize}

These attributes are difficult to quantify in the absolute sense \cite{avizienis2001fundamental}. Real systems can never be totally available, reliable or safe: treats are inevitable in real systems. In CPS paradigm, typically we consider two different types of threats, including \emph{random} faults and failures, and \emph{strategic} threats consisting of attacks by an adversary with an objective to maximally disrupt the CPS operations. 
The development of a dependable computing system calls for the combined utilization of a set of methods and techniques which can provide threat prevention, threat tolerance, threat removal and threat forecasting. The concept of dependability must be explored in terms of threats to dependability and means to attain it.

In order for a system to be dependable, it must support the following
:

\begin{itemize}
    \item \textit{Threat prevention}: how to prevent the occurrence or introduction of threats; 
    \item \textit{Threat tolerance}: how to deliver correct service in the presence of threats;
    \item \textit{Threat removal}: how to reduce the number or severity of threats; 
    \item \textit{Threat forecasting}: how to estimate the present number, the future incidence, and the likely consequences of threats.
\end{itemize}

Embedded systems electronics, in general, are far more predictable and reliable than general-purpose computing\cite{lee2008cyber}. CPS should increase the reliability of embedded systems. Reliability and predictability of CPS are mandatory for their deployment in critical applications such as healthcare, air traffic control and automotive safety \cite{lee2015past}. Other attributes like security must also be dealt with. Due to the growing level of integration of new information technologies, the modern CPS face uncertainties both from physical world and cyber-components of the system \cite{zhu2015game}. These vulnerabilities in the CPS can expose the system to various potential threats and risks from attackers which can lead to intensive damages. Hence, it is imperative to consider both cyber and physical uncertainties in designing reliable and robust CPS.


The robustness of CPS is its ability to resist a known range of uncertain disturbances and parameters, while its security describes the ability to withstand and be protected from unanticipated and malicious events. These two properties are pre-event, that is, the CPS is designed to be secure and robust. Despite many efforts, the designing of robust and secure systems is very costly and impossible to achieve perfect security and robustness \cite{zhu2015game}. Consequently, it becomes necessary to analyze the resilience of the system (post-event), which is the system's ability to recover after the occurrence of disruptive events.

The concept of \textit{security} comes in handy when addressing the dependability of computing or communication systems. Security has been recognized as a composite confidentiality, integrity, and availability \cite{avizienis2004basic}. Confidentiality (trust that information will not be disclosed without authorization) is another concept that has gained prominence in the context of security. Figure \ref{Dep_Sec} depicts the relationship between dependability and security in terms of the principal attributes of dependability. The development of a resilient CPS requires a deep understanding of disruptions caused by cyber attacks. This requires an evaluation of CPS dependability on its cyber infrastructure and its ability to tolerate the failures events \cite{sridhar2012cyber}.

CPS are complex systems and have many loops of operation working at different scales of time and space \cite{sha2008cyber}. The reliability of a complete system can be determined from the reliability of its components. The probability of failure for a system without redundancy is more than the probability of failure of any of its components. A CPS's properties depend on both component properties and the system architecture \cite{sha2008cyber}. Reliability and dependability analysis of CPS is usually based on traditional techniques for systems reliability analysis \cite{rausand2004system}. Some contributions in reliability analysis of CPS include \cite{mitchell2016modeling}, \cite{liutransportation},\cite{davis2015cyber} and \cite{engel2011unreliable}. Comprehensive research on the dependability of CPS is still needed to predict their reliability and formulate methods to improve dependability.

\textit{Reliability analysis} allows us to identify problems in telecommunication networks as well as to determine the particular redundancy requirement of a particular network \cite{ahmed2017reliability}. \textit{Reliability modeling} comes before analysis in the design phase. This is followed by reliability analysis in later design stages when more precise implementation details are available \cite{bernardi2013model}. Reliability modeling is the development of a  model to predict the reliability or vulnerability of a system from information available. Reliability modeling allows us to calculate dependability metrics for a system. It can be achieved by combinatorial models such as Reliability Block Diagram (RBD) and Fault Tree (FT) or through state-based stochastic models such as Markov Chains (MC) and Stochastic PetriNets (SPN) \cite{fernandes2012dependability}. Combinatorial models allow representation of system reliability in terms of reliability of components and provide closed form equations. However, they cannot represent failure dependencies and resource constraints that are required for maintenance policies and describing redundant mechanisms \cite{ebeling2004introduction}. State-based modeling, which predicts reliability analytically, can represent complex redundant mechanisms. They can also be used to predict maintenance policies \cite{fernandes2012dependability}. However, the possibility of state-space explosion must be dealt with \cite{weber2012overview}. More recently Bayesian Networks (BNs) have been employed for reliability modeling, either directly or by mapping fault trees into them \cite{bobbio2001improving,weber2012overview}.  BNs are a graphical representation of conditional dependencies of system components and take the form of a Directed Acyclic Graph (DAG). They represent component interactions in a probabilistic way. Petri Nets (PNs) and SPNs are a form of BNs which allow us to model the dynamic (temporal and cause and effect) behavior of network components more effectively. They are particularly suited to modeling state transitions and information flow in complex systems \cite{knezevic2001reliability}. They allow numerical analysis as well as stochastic simulation.    

Also known as a Dependence Diagram (DD), a reliability block diagram is a series and parallel arrangement of blocks that represent the probability of failure of a system in terms of reliability of its components (blocks).  A system represented by a RBD will work only if there is at least one series path of working blocks along the span of the diagram. RBDs are intended for systems without the ability to repair and where the order of failures does not matter. FT diagrams are a visual representation of the logical relationship between component or sub-system failures. A basic event in FTA is a top event of a fault tree that represents a system event of interest that is connected by logical gates to component failures \cite{xing2008fault}. FTs and RBDs reveal how individual component failures contributes towards the failure of a whole system. FTs can be particularly useful for identifying critical components. RBDs and FTs are combinatorial methods in that they allow us to learn how a combination of certain events can trigger another event. They do not take into account the order of events \cite{dutuit1997dependability}. Markov Chain based modeling is suitable when the order of failure matters or when repairs are possible. FTs and RBDs are used to model reliability and estimate availability in both early and later stages of the design. Models based on Markov chains are generally used in later design phases to evaluate or compare different design alternatives  \cite{ahmed2017reliability}.

Models developed using these or similar techniques can be analyzed using traditional analytical methods or through simulation tools. Formal methods are now gaining attention as a useful tool for modeling reliability and validating models \cite{ahmed2017reliability}. Analytical models rely on the abstraction, simplification and unrealistic assumptions of the complex system. This can make them error prone, particularly in large complex systems. Formal methods are a rigorous method for analysis compared to traditional analytic and simulation techniques. Reliability assessment, analysis and modeling of networks are beyond the scope of this paper. The reader can find a comprehensive study on reliability analysis in a paper by Ahmed et al. \cite{ahmed2017reliability}.

These classical reasoning and modeling techniques are based on Boolean logic, analytical models, determinism and crisp classification. In the realm of modeling the system (or CPS) is supposed to have the complete and precise information needed to solve the particular problem. In the real world, relevant information is often available in the form of empirically acquired prior knowledge and system behavior determined from past input-output data. In many instances, multiple solutions may exist within a large scale solution space that can fit our problem. Soft computing technologies encompass a set of flexible computing tools that can deal with imprecise information and search for approximate answers \cite{bonissone1997soft}. Multiple soft computing techniques can be used in cyber-physical and other complex systems to improve system dependability or to model dependability. Unlike sensor networks, CPS perform physical actions that are characterized by distributed control loops which receive essential feedback from the environment. In addition, the number of nodes and communication capabilities in CPS vary significantly. Such ecosystem of complex smart systems leads to a hybrid system which makes use of fuzzy sets, neural networks and evolutionary computation in different stages or processes \cite{juuso2004integration}.




\section{Soft Computing for Dependable CPS}

\begin{figure*}[!ht]
\centering
\includegraphics[width=7.0in]{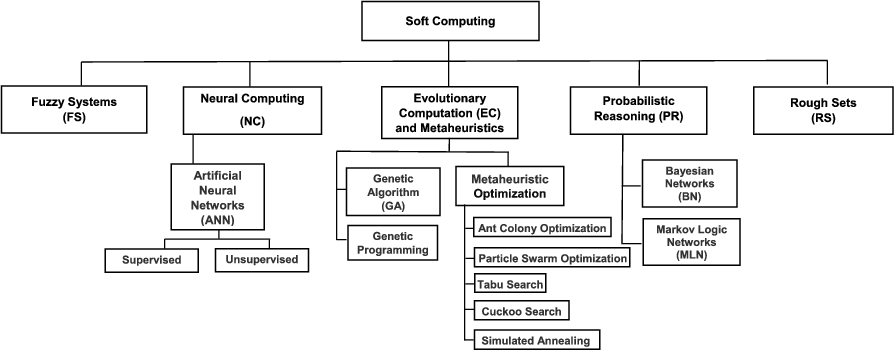} 
\caption{Taxonomy of Soft Computing (adapted from \cite{chaturvedi2008soft}\cite{maddalena2012handbook} \cite{bello2012rough})}
\label{Tex}
\end{figure*}

Soft computing is a collection of computing methodologies that include Fuzzy Logic (FL), (Artificial) Neural Networks (ANN), Evolutionary Computation (EC) as their principal members \cite{zadeh1994soft}. The taxonomy of soft computing techniques is shown in Figure \ref{Tex}. These methodologies are complementary and symbiotic for the most part as evident from the use of a combination of these methodologies in intelligent systems \cite{zadeh1994soft}. Later Probabilistic Reasoning (PR), Machine Learning (ML), Belief Networks (i.e., Bayesian Networks (BNs)), Chaos Theory, parts of Learning Theory and Wisdom-based Expert Systems were subsumed under the same umbrella \cite{chaturvedi2008soft}. Rough Sets (RS) are also considered by many as a soft computing technique \cite{maddalena2012handbook,bello2012rough}.

These soft computing techniques have been being used to improve the dependability aspects like reliability or security of complex systems. They have also been used in modeling the reliability of complex systems and computer networks. They are required in instances when it is hard to obtain an analytical model to evaluate system reliability \cite{ebrahimipour2013emotional} and also prove useful when Monte Carlo simulations are not feasible to evaluate reliability. Soft computing techniques can be a substitute for simulation models (as meta-models) \cite{ebrahimipour2013emotional}. They are also useful in solving complex optimization problems, particularly when information is vague or incomplete.  The strengths and weakness of different soft computing techniques is listed in table \ref{table: Su}. 

In this section, a summary of applications of soft computing techniques for CPS dependability is provided. While this work is focused on CPS, we have allowed inclusion of works on the dependability of software systems, complex systems, computer networks and similar components of broad CPS.

\begin{table*}[!ht]
\centering
\scriptsize
\caption{Strengths and weaknesses of various of soft computing techniques (derived from  \cite{chaturvedi2008soft,maddalena2012handbook,aliev2012soft,frias2005modeling})}
\begin{tabular}{ |m{3.4cm} | m{1.4cm} |m{1.8cm} |m{2.0cm} |m{2.0cm} |m{1.4cm}| }
\hline
\textbf{Feature}&
\textbf{Fuzzy Logic (FL)}
& \textbf{Artificial Neural Nets (ANN)}
& \textbf{Evolutionary Computation (EC)}  
& \textbf{Probabilistic Reasoning (PR)}
& \textbf{Rough Sets (RS)}
 \\ \hline
Training through data&No&Yes&Yes&No&Yes\\
\hline
Parallel processing ability&No&Yes&Yes&No&No\\
\hline
Symbolic  input required &Yes&Yes&Yes&Yes&Yes\\
\hline
Unlabeled data support&NA&Yes&Yes&No&Yes\\
\hline
Computational complexity&Low&High&High&Medium&Low\\
\hline 
Incomplete information support&Yes&No&No&No&Yes\\
\hline
Linguistic information support&Yes&No&No&No&No\\ 
\hline

\end{tabular} 
 
\label{table: Su}
\end{table*}

\subsection{Fuzzy Set Theory and Fuzzy Logic}
Fuzzy set theory has been incorporated into reliability theory by modifying the conventional assumptions about the reliability of a component/system, i.e., binary state (success or failure) and probability measure of its reliability \cite{ravi2000fuzzy}.
Fuzzy Logic (FL) was designed to handle imprecision using approximate reasoning \cite{zadeh1994soft}. It is a pioneering technology in granular computing. It has been described as a form of computing with words \cite{zadeh1998some} since it mimics the human method of reasoning with words by using linguistic variables and values. FL is a generalization of Boolean logic \cite{chaturvedi2008soft} centered on fuzzy sets. Any object belonging to a fuzzy set can have a degree of membership (a real number between 0 and 1) for that particular set. Fuzzy inference maps inputs to outputs using FL. This mapping can then be used to infer patterns or make decisions. This inference involves four steps, namely \textit{fuzzification} (real value to fuzzy membership values), \textit{rule evaluation}, \textit{aggregation} (of rules) and finally \textit{defuzzification} \cite{negnevitsky2005artificial}. Fuzzy inference is relatively simple to implement and finds extensive use in contemporary control systems for consumer appliances.

FL has been used in the analysis of structural reliability, fault detection, probist systems, software reliability, safety, security and risk engineering \cite{cai1996system}. FL was traditionally focused on reliability analysis of components or systems. However, there are few cases where fuzzy set theory has been used for global optimization of reliability \cite{ravi2000fuzzy}. Mahapatra et al. have discussed the optimization of reliability for series and complex systems with (conflicting) reliability and cost objectives, using fuzzy multi-objective optimization method with fuzzy parameters \cite{mahapatra2014reliability}. In another paper, the same authors have used intuitionistic fuzzy optimization for the reliability of complex systems \cite{mahapatra2010intuitionistic}.

Traditional reliability modeling techniques are based on statistics of past performance of a system or components. Sometimes it is not feasible to obtain such long-term data accurately. Classical reliability treatment also involves human judgment to some extent \cite{onisawa1990application}. Fuzzy probabilities or possibilities \cite{kaufmann1975introduction} provide a flexible and efficient means for modeling such systems \cite{knezevic2001reliability}. 
A system success and failure can be represented by fuzzy states, and systems can be in one of these two states to some extent. Further, the failure behavior of the systems can be fully characterized by the possibility measures instead of probabilities. Fuzzy logic and possibility theory are an alternative to probabilistic modeling \cite{zadeh1978fuzzy}. Probability is the degree of likelihood assumed from the frequency of occurrence of an event \cite{keller1989further}. The possibility is the degree of feasibility or ease of attainment \cite{zadeh1999fuzzy}. In practice, it makes more sense to use possibility, particularly in the design phase when actual frequency tables of a component's reliability are not available \cite{keller1989further}. The probability assumption is also not valid in the case of the small size of samples \cite{kai1991fuzzy}.   

\begin{table*}
\centering
\scriptsize
\caption{Applications of Fuzzy Logic in Improving System Performance and System Dependability Modeling of CPS}
\begin{tabular}{ |m{2.6cm}|m{1.8cm}|m{6cm}|m{2.6cm}|}
\hline
\textbf{Paper Reference}
& \textbf{Soft Computing Technique}
& \textbf{Description}
& \textbf{Dependability Attribute}
 \\ \hline
\begin{tabular}[c]{@{}l@{}}Huiling et al. 2008 \cite{huiling2008software} \end{tabular}   
& \begin{tabular}[c]{@{}l@{}}FL\end{tabular}  
& \begin{tabular}[c]{@{}l@{}}Software dependability evaluation. Fuzzy inference of \\dependability attributes
\end{tabular}  
& \begin{tabular}[c]{@{}l@{}}Availability, Reliability,\\ Safety, Confidentiality, \\Integrity,
Maintainability\end{tabular}  
\\\hline
\begin{tabular}[c]{@{}l@{}}Toosi et al. 2007 \cite{toosi2007new}\end{tabular}   
& \begin{tabular}[c]{@{}l@{}}FL, ANN, GAs \end{tabular} 
& \begin{tabular}[c]{@{}l@{}}Neuro Fuzzy classifiers for an intrusion detection system\\ for computer  networks (and GA to optimize structure of\\ fuzzy decision engine). FL and ANN used to identify\\ intrusion.
\end{tabular}  
&\begin{tabular}[c]{@{}l@{}}Security,  Confidentiality\end{tabular}  
\\\hline
\begin{tabular}[c]{@{}l@{}}Knezevic et al. 2001 \cite{knezevic2001reliability} \end{tabular}   
&\begin{tabular}[c]{@{}l@{}}FL \end{tabular}  
&\begin{tabular}[c]{@{}l@{}}Reliability modeling using SPNs and calculation \\of reliability indices using FL and lambda-tau technique
\end{tabular}  
&\begin{tabular}[c]{@{}l@{}}Reliability, Availability,\\Maintainability\\\end{tabular}  
\\\hline
\begin{tabular}[c]{@{}l@{}}Garg et al. 2014  \cite{garg2014approach}\end{tabular}   
&\begin{tabular}[c]{@{}l@{}}FL \end{tabular}  
&\begin{tabular}[c]{@{}l@{}}Estimation of reliability indices for industrial systems\\using Lambda–Tau method supported by FL (and ABC \\algorithm for finding fuzzy membership)
\end{tabular}  
&\begin{tabular}[c]{@{}l@{}}Reliability, Availability\end{tabular}  
\\\hline
\begin{tabular}[c]{@{}l@{}}Rotshtein et al. 2012  \cite{rotshtein2012reliability}\end{tabular}   
&\begin{tabular}[c]{@{}l@{}}FL \end{tabular}  
&\begin{tabular}[c]{@{}l@{}}Modeling and optimization of reliability using FL \\and chaos theory
\end{tabular}  
& \begin{tabular}[c]{@{}l@{}}Reliability\\ \end{tabular}
\\\hline
\begin{tabular}[c]{@{}l@{}}Mahapatra  et al. 2006 \cite{mahapatra2006fuzzy}\end{tabular}  
&\begin{tabular}[c]{@{}l@{}}FL \end{tabular}  
&\begin{tabular}[c]{@{}l@{}}Fuzzy multi-objective optimization method for\\ reliability optimization of series and complex systems\end{tabular}  
& \begin{tabular}[c]{@{}l@{}}Reliability, Availability\end{tabular}
\\\hline
\begin{tabular}[c]{@{}l@{}}Cho et al. 2002 \cite{cho2002incorporating}\end{tabular}   
&\begin{tabular}[c]{@{}l@{}}FL, ANN\end{tabular}  
&\begin{tabular}[c]{@{}l@{}}Intrusion and anomaly detection in computing systems\\ using HMM and ANN\end{tabular}  &\begin{tabular}[c]{@{}l@{}}Security, Confidentiality\end{tabular}
\\\hline
\begin{tabular}[c]{@{}l@{}}Pandey et al. 2009 \cite{pandey2009fuzzy}\end{tabular}   
&\begin{tabular}[c]{@{}l@{}}FL\end{tabular}  
&\begin{tabular}[c]{@{}l@{}}Early Software Fault Prediction (based on reliability\\ metrics and expert knowledge)
\end{tabular}  
&\begin{tabular}[c]{@{}l@{}}Reliability, Availability,\\Maintainability\end{tabular}
\\\hline
\begin{tabular}[c]{@{}l@{}}Mahapatra et al. \\2014 \cite{mahapatra2014reliability}\end{tabular}   
&\begin{tabular}[c]{@{}l@{}}FL\end{tabular}  
&\begin{tabular}[c]{@{}l@{}}Complex system reliability optimization using intuitionistic\\ fuzzy sets
\end{tabular}  
&\begin{tabular}[c]{@{}l@{}}Reliability\end{tabular}
\\\hline
\begin{tabular}[c]{@{}l@{}}Ebrahimipour et al. \\2013 \cite{ebrahimipour2013emotional}\end{tabular}   
&\begin{tabular}[c]{@{}l@{}}FL\end{tabular}  
&\begin{tabular}[c]{@{}l@{}}Emotional learning-based fuzzy inference system \\to improve performance of reliability evaluation systems.\\ANN and GA etc. used as system reliability meta-models\\hybridized using meta-heuristics. 
\end{tabular}  
&\begin{tabular}[c]{@{}l@{}}Reliability\end{tabular}
\\\hline
\begin{tabular}[c]{@{}l@{}}Tyagi  et al. 2014 \cite{tyagi2014adaptive}\end{tabular}   
&\begin{tabular}[c]{@{}l@{}}FL, ANN\end{tabular} 
&\begin{tabular}[c]{@{}l@{}}Estimating reliability of component-based software\\systems
\end{tabular}  
&\begin{tabular}[c]{@{}l@{}}Reliability, Availability\end{tabular}
\\\hline
\end{tabular}

\label{table:FL}
\end{table*}
Fuzzy logic has been used to model the reliability of software systems. Cai et al. \cite{cai1991critical} have discussed fuzzy software reliability models as a substitute for probabilistic models. Khatatneh et al. \cite{khatatneh2009software} have developed a fuzzy logic based model to predict software reliability. Researchers have used fuzzy theory in conjunction with fault tree analysis \cite{tanaka1983fault,singer1990fuzzy} and in fuzzy Lambda-Tau methodology \cite{knezevic2001reliability}. Cheng et al. \cite{cheng1993fuzzy} proposed a fuzzy number over the interval of confidence as an alternative to probability in reliability analysis. Huiling et al. \cite{huiling2008software} have presented a fuzzy inference based system to evaluate all software dependability in terms of its attributes. Their proposed method consists of a dependability evaluation indicator system and a fuzzy classifier with a feedback system that reconfigures itself with training from expert's opinion and quantitative data.
 
In some cases, statistical data on system reliability may not contain information about causes of failures which is not suitable for reliability modeling. Rotshtein et al. \cite{rotshtein2012reliability} have discussed a reliability evaluation method based on the integration of the fuzzy modeling of reliability with the technique of time series by using chaos theory for analysis of nonlinear time series. Their method is capable of reflecting system reliability dynamics. Mahapatra et al. have presented a fuzzy logic based technique for finding the optimum system reliability of complex systems, constrained by a system cost \cite{mahapatra2014reliability}. Their system uses Intuitionistic Fuzzy Set (IFS) which is a generalization of fuzzy set theory designed to deal with the vagueness and imprecision of data. IFS has been used to model human decision making \cite{atanassov1986intuitionistic}. The model presented by Mahapatra et al. \cite{mahapatra2014reliability} trades off some precision in reliability optimization for system efficiency. In an earlier paper, Mahapatra  et al. \cite{mahapatra2006fuzzy} have discussed reliability apportionment for complex systems in the presence imprecisely known costs of components. They have used fuzzy set theory to handle imprecise data and multi-objective programming using Fuzzy Non-Linear Programming (FNLP) with fuzzy parameters. Ebrahimipour et al. \cite{ebrahimipour2013emotional} have used FL in conjunction with ANNs for solving the Redundancy Allocation Problem (RAP) under constraints. RAP aims to maximize system reliability during the design phase. Pandey et al. \cite{pandey2009fuzzy} have presented a fuzzy logic based inference model to predict software faults. Their system requires software reliability metrics and a model based on the developer's capability maturity along with expert's opinions (subjective knowledge). 

Fuzzy Logic is a robust and relatively simpler soft computing technique for classification. In many instances, determination of fuzzy membership functions required in FISs, is performed by other techniques. Huang et al. \cite{huang2006bayesian} have used GAs to estimate boundary values of the fuzzy membership functions, and ANNs to estimate fuzzy parameters for their Bayesian model for reliability analysis. Toosi et al. \cite{toosi2007new} have discussed an Intrusion Detection System (IDS) built upon a FL aided by ANNs. They have used GAs to optimize parameters for their fuzzy classifier. Knezevic  et al. \cite{knezevic2001reliability} have used Lambda-Tau method with the aid of fuzzy logic to calculate reliability indices like availability, Mean Time To Failure (MTTF), Mean Time To Recovery (MTTR) etc. They have used fuzzy arithmetic with PNs to model reliability with the benefit of increased flexibility and requirement of a smaller data set of prior reliability. Garg et al. \cite{garg2014approach} have presented a similar method to calculate reliability indices for industrial systems using Lambda–Tau technique with FL and artificial bee colony algorithm to calculate fuzzy membership degrees. Tyagi et al. \cite{tyagi2014adaptive} have calculated the reliability of  component based software systems using an adaptive neuro-fuzzy inference system (ANFIS).

FL is also used in conjunction with or to aid other techniques for reliability modeling improvement or optimization. Lin et al. \cite{lin2006self} have used FL to tune the probabilities of their genetic operators in their GA for reliable communication network design. FL is used as a classifier in the IDS by Cho et al. \cite{cho2002incorporating} for computer networks.
Application of Fuzzy Set Theory and FL in CPS reliability analysis and improvement is summarized in Table \ref{table:FL}. It can be seen from the Table \ref{table:FL} that FL has mainly been used in fault diagnosis, RAP, software reliability evaluation, safety and security assessment and intrusion detection \cite{cai1996system}. Research into the use FL in dependability engineering has primarily focused on reliability analysis of systems. However, there are a few cases where FST has been used for global optimization of reliability \cite{ravi2000fuzzy}.

\subsection{Evolutionary Computation and Meta-Heuristics}

Evolutionary Computation is a mechanism for systematic random search directed at finding an optimum solution to a problem \cite{zadeh1998some}. Genetic Algorithms (GA) and other modes of genetic computing are special cases of EC. GA generate a population of candidate solutions to a particular problem and evaluate them based on a fitness function, and select good solutions. Similar to natural evolution, surviving solutions retain the fittest parts from previous generations \cite{goldberg2006genetic}. The best solution in each population usually survives as an elite individual and passes its characteristics to its offspring. Genetic programming (GP) is an extension of genetic algorithms. It is a technique to encode computer programs as a set of genes that may evolve using an evolutionary algorithm. EC techniques also include metaheuristic population-based optimization algorithms with names inspired by nature. Particle Swarm Optimization (PSO), Ant Colony Optimization (ACO) and Cuckoo Search (CS) \cite{yang2009cuckoo} are some prominent EC algorithms. Metaheuristic optimization algorithms like Tabu Search (TS) \cite{glover1997tabu}, and Simulated Annealing (SA) (a stochastic optimization metaheuristic) \cite{myllymaki1999massively}, may also be categorized in the same group as EC.

EC has seen a rapid growth in terms of applications for CPS reliability. GA are a family of heuristic optimization techniques and used to find optimal solutions to diverse problems. However, optimality is not guaranteed. Because GA's ability to dig up good solutions mostly depends upon proper customization of the fitness functions, encoding, and breeding operators for the specific problem \cite{coit1996reliability}. Optimization approaches like Dynamic Programming (DP), integer programming, Mixed Integer Non-Linear Programming (MINLP), and other heuristics are used to determine optimal solutions. GAs have solved many difficult engineering problems. They are suited to solve combinatorial optimization problems within complex search spaces. However, there are relatively few examples of their use in the field of reliability analysis. One of the first proposals for designing an optimization tool to maintain scheduling activities using GA was presented by Munoz et al. \cite{munoz1997genetic}. Lapa et al. \cite{lapa2000maximization} applied a new approach using GA to optimize maintenance and inspection  intervals. In this paper, the authors tried to search optimal times for performing preventive maintenance interventions. Yang et al. \cite{yang2000optimization} suggested a plant level surveillance policy optimization technique. A similar approach was investigated by Lapa et al. \cite{lapa2002application,lapa2006model}. 


\begin{table*}[!ht]
\centering
\scriptsize
\caption{Applications of Evolutionary Computation (EC) and Metaheuristics for Improving System Performance and System Dependability Modeling of CPS}
\begin{tabular}{ |m{2.6cm}|m{1.7cm}|m{6.2cm}|m{2.8cm}|}
\hline
\textbf{Paper Reference}
& \textbf{Soft computing Technique}
& \textbf{Description}
& \textbf{Dependability Attribute}
\\\hline
\begin{tabular}[c]{@{}l@{}}Echtle et al. 2003\cite{echtle2003genetic}\end{tabular}   
&\begin{tabular}[c]{@{}l@{}}GA\end{tabular}  
&\begin{tabular}[c]{@{}l@{}}Estimating reliability of component-based software systems.\\Custom fitness function similar to reachability analysis used\end{tabular}  
& \begin{tabular}[c]{@{}l@{}}Reliability, Availability,\\Maintainability\end{tabular}
\\\hline
\begin{tabular}[c]{@{}l@{}}Coit et al. 1996  \cite{coit1996reliability}\end{tabular}   
&\begin{tabular}[c]{@{}l@{}}GA\end{tabular}  
&\begin{tabular}[c]{@{}l@{}}GA for allocation of redundancy in series-parallel systems
\end{tabular}  
&\begin{tabular}[c]{@{}l@{}}Reliability, Availability\end{tabular}
\\\hline
\begin{tabular}[c]{@{}l@{}}Lapa et al. 2006 \cite{lapa2006model}\end{tabular}   
&\begin{tabular}[c]{@{}l@{}}GA\end{tabular}  
&\begin{tabular}[c]{@{}l@{}}Use of Genetic Algorithms for searching optimum preventive\\ maintenance policies based on constraints including the \\repair of cost and outage time. FTA, MC, min cut sets, etc,\\ used for dependability modeling
\end{tabular}  
&\begin{tabular}[c]{@{}l@{}}Reliability, Availability,\\Maintainability\end{tabular}
\\\hline 
\begin{tabular}[c]{@{}l@{}}Duan et al. 2015 \cite{duan2015reconfiguration}\end{tabular}   
&\begin{tabular}[c]{@{}l@{}}GA\end{tabular}  
&\begin{tabular}[c]{@{}l@{}}Power distribution network reconfiguration method for loss \\reduction and improved reliability 
\end{tabular}  
& \begin{tabular}[c]{@{}l@{}}Reliability, Availability\end{tabular}
\\\hline
\begin{tabular}[c]{@{}l@{}}Tian et al. 2005 \cite{tian2005line}\end{tabular}   
&\begin{tabular}[c]{@{}l@{}}GA, ANN\end{tabular}  
&\begin{tabular}[c]{@{}l@{}}Adaptive on-line modeling of software reliability prediction\\through ``evolutionary connectionist'' approach. Dependability\\ modeling through Bayesian regularization and\\ ANN+Levenberge-Marquardt algorithm
\end{tabular}  
& \begin{tabular}[c]{@{}l@{}}Reliability, Availability\end{tabular}
\\\hline

\begin{tabular}[c]{@{}l@{}}Tian et al. 2005 \cite{tian2005evolutionary}\end{tabular}   
&\begin{tabular}[c]{@{}l@{}}GA, ANN\end{tabular}  
&\begin{tabular}[c]{@{}l@{}}Modeling of Software failure time prediction.\\ANN+Levenberge-Marquardt algorithm with Bayesian \\regularization for modeling dependability.
\end{tabular}  
& \begin{tabular}[c]{@{}l@{}}Reliability, Availability\end{tabular}
\\\hline
\begin{tabular}[c]{@{}l@{}}Zhao and Liu 2003 \cite{zhao2003stochastic}\end{tabular}   
&\begin{tabular}[c]{@{}l@{}}GA, ANN\end{tabular}  
&\begin{tabular}[c]{@{}l@{}}Stochastic Simulation, GA and ANN for solving general\\resource allocation problem (RAP)
\end{tabular}  
& \begin{tabular}[c]{@{}l@{}}Reliability, Availability\end{tabular}
\\\hline
\begin{tabular}[c]{@{}l@{}}Aljahdali  et al. 2009 \cite{aljahdali2009software}\end{tabular}
&\begin{tabular}[c]{@{}l@{}}GA\end{tabular}
&\begin{tabular}[c]{@{}l@{}}Ensemble models trained though GA to predict software\\reliability\end{tabular}
& \begin{tabular}[c]{@{}l@{}}Reliability, Availability\end{tabular}
\\\hline
\begin{tabular}[c]{@{}l@{}}Elkoujok et al. 2013 \cite{elkoujok2013application}\end{tabular}
&\begin{tabular}[c]{@{}l@{}}GA\end{tabular}
&\begin{tabular}[c]{@{}l@{}}Analytically redundancy approach to detect and isolate\\sensor faults in non-linear systems. GA and evolving\\ Takagi-Sugeno algorithm used for dependability modeling.\end{tabular}
& \begin{tabular}[c]{@{}l@{}}Reliability, Maintainability\end{tabular}
\\\hline
\begin{tabular}[c]{@{}l@{}}Lin et al. 2006 \cite{lin2006self}\end{tabular}
&\begin{tabular}[c]{@{}l@{}}GA, FL\end{tabular}
&\begin{tabular}[c]{@{}l@{}}(self-controlled) GA for designing reliable communication\\networks. Custom graph used for communication network\\dependability modeling\end{tabular}
&\begin{tabular}[c]{@{}l@{}}Reliability, Availability \end{tabular}
\\\hline
\begin{tabular}[c]{@{}l@{}}Mitra et al. 2009 \cite{mitra2009real}\end{tabular}
&\begin{tabular}[c]{@{}l@{}}PSO\end{tabular}
&\begin{tabular}[c]{@{}l@{}}An intelligent dynamic generator and load reconfiguration \\strategy for an electric ship power system\end{tabular}
&\begin{tabular}[c]{@{}l@{}}Availability, Maintainability\end{tabular}
\\\hline
\begin{tabular}[c]{@{}l@{}}Robinson et al. 2005 \cite{robinson2005reliability} \end{tabular}
&\begin{tabular}[c]{@{}l@{}}PSO\end{tabular}
&\begin{tabular}[c]{@{}l@{}}Reliability analysis of bulk power systems (electrical grids).\\PSO in identifying critical elements\end{tabular}
&\begin{tabular}[c]{@{}l@{}}Reliability, Availability,\\ Maintainability\end{tabular}
\\\hline
\begin{tabular}[c]{@{}l@{}}Bashir et al. 2009\cite{bashir2009applying}\end{tabular}
&\begin{tabular}[c]{@{}l@{}}PSO, ANN\end{tabular}
&\begin{tabular}[c]{@{}l@{}}Predicting hourly electric load demand using adaptive ANNs\\ with assistance from PSO for calculating weights for ANN\end{tabular}
&\begin{tabular}[c]{@{}l@{}}Availability, Maintainability\end{tabular}
\\\hline
\begin{tabular}[c]{@{}l@{}}Khan et al. 2014 \cite{khan2014computational}\end{tabular}
&\begin{tabular}[c]{@{}l@{}}PSO\end{tabular}
&\begin{tabular}[c]{@{}l@{}}Fault tolerant autonomous control of aircraft CPS\end{tabular}
&\begin{tabular}[c]{@{}l@{}}Reliability, Maintainability\end{tabular}
\\\hline
\begin{tabular}[c]{@{}l@{}}Liang et al. 2004 \cite{liang2004ant}\end{tabular}
&\begin{tabular}[c]{@{}l@{}}ACO\end{tabular}
&\begin{tabular}[c]{@{}l@{}}Redundancy allocation problem (RAP) using ant colony \\optimization (ACO)\end{tabular}
&\begin{tabular}[c]{@{}l@{}}Reliability, Availability\end{tabular}
\\\hline
\begin{tabular}[c]{@{}l@{}}Zhao et al. 2007\cite{zhao2007reliability}\end{tabular}
&\begin{tabular}[c]{@{}l@{}}ACO\end{tabular}
&\begin{tabular}[c]{@{}l@{}}Multi-objective ACO to solve reliability optimization\\for series-parallel systems\end{tabular}
&\begin{tabular}[c]{@{}l@{}}Reliability, Availability\end{tabular}
\\\hline
\begin{tabular}[c]{@{}l@{}}Caserta et al. 2009\cite{caserta2009tabu}\end{tabular}
&\begin{tabular}[c]{@{}l@{}}Tabu Search\end{tabular}
&\begin{tabular}[c]{@{}l@{}}Design of reliable software systems with optimization of\\ redundancy\end{tabular}
&\begin{tabular}[c]{@{}l@{}}Reliability, Availability,\\Maintainability \end{tabular}
\\\hline
\begin{tabular}[c]{@{}l@{}}Kulturel et al. 2003 \cite{kulturel2003efficiently}\end{tabular}
&\begin{tabular}[c]{@{}l@{}}Tabu search\end{tabular}
&\begin{tabular}[c]{@{}l@{}}TS as an efficient alliterative to GAs for RAP\end{tabular}
&\begin{tabular}[c]{@{}l@{}}Reliability, Availability,\\Maintainability\end{tabular}
\\\hline
\begin{tabular}[c]{@{}l@{}}Ramirez et al. 2006 \cite{ramirez2006new}\end{tabular}
&\begin{tabular}[c]{@{}l@{}}Tabu search\end{tabular}
&\begin{tabular}[c]{@{}l@{}}Optimal planning of power distribution systems using\\Tabu search and FL \end{tabular}
&\begin{tabular}[c]{@{}l@{}}Reliability, Availability,\\Maintainability\end{tabular}
\\\hline
\begin{tabular}[c]{@{}l@{}}Pierre et al. 1997 \cite{pierre1997tabu}\end{tabular}
&\begin{tabular}[c]{@{}l@{}}Tabu search\end{tabular}
&\begin{tabular}[c]{@{}l@{}}Network reliability and redundancy  allocation\end{tabular}
&\begin{tabular}[c]{@{}l@{}}Reliability, Availability,\\Maintainability\end{tabular} 
\\\hline
\begin{tabular}[c]{@{}l@{}}Valian et al. 2013 \cite{valian2013improved}\end{tabular}
&\begin{tabular}[c]{@{}l@{}}Cuckoo search\end{tabular}
&\begin{tabular}[c]{@{}l@{}}Reliability optimization and redundancy  allocation\end{tabular}
&\begin{tabular}[c]{@{}l@{}}Reliability, Availability\\ Maintainability\end{tabular}
\\\hline
\begin{tabular}[c]{@{}l@{}}Teske  et al. 2015 \cite{teske2015efficient}\end{tabular}
&\begin{tabular}[c]{@{}l@{}}Cuckoo search\end{tabular}
&\begin{tabular}[c]{@{}l@{}}Fault detection in parallel and  distributed systems\end{tabular}
&\begin{tabular}[c]{@{}l@{}}Reliability,\\Maintainability\end{tabular}
\\\hline

\begin{tabular}[c]{@{}l@{}}Pai et al. 2006 \cite{pai2006software}\end{tabular}
&\begin{tabular}[c]{@{}l@{}}SA\end{tabular}
&\begin{tabular}[c]{@{}l@{}}Software reliability prediction. SA, SVM used for \\dependability modeling \end{tabular}
&\begin{tabular}[c]{@{}l@{}}Reliability, Availability\end{tabular}
\\\hline

\begin{tabular}[c]{@{}l@{}}Attiya et al. 2006 \cite{attiya2006task}\end{tabular}
&\begin{tabular}[c]{@{}l@{}}SA\end{tabular}
&\begin{tabular}[c]{@{}l@{}}System reliability optimization through task allocation\\in distributed systems\end{tabular}
&\begin{tabular}[c]{@{}l@{}}Reliability\end{tabular}
\\\hline

\end{tabular}

\label{table:GA}
\end{table*}

Painton and Campbell \cite{painton1993identification,painton1995genetic} used GA to find maximum reliability solutions to satisfy specific cost constraints for designs with fixed configuration and predetermined incremental decreases in component failure rates.  Their flexible algorithm can optimize either availability or Mean Time Between Failures (MTBF).  Ida et al. \cite{ida1994system} applied a GA to find solutions to a RAP in the presence of multiple failure modes. GA have also been used in optimization of series parallel systems \cite{hsieh1998genetic}. Zhao and Liu \cite{zhao2003stochastic} have proposed a solution to the general redundancy allocation problem using GA and ANNs. Coit and Smith \cite{coit1994use} analyzed a complex series-parallel system (eight subsystems and ten unique component choices per subsystem). Their search space had more than 1030 solutions but the GA could converge to a solution after analyzing less than 10-24\%  of the search space. Ecthle et al. have presented a GA and a special fitness function for automated system design of fault tolerant structures \cite{echtle2003genetic}. Their algorithm finds a redundant arrangement of components. Duan et al. \cite{duan2015reconfiguration} have presented an enhanced GA for network reconfiguration for loss reduction and improved reliability in smart grids. Elkoujok et al. \cite{elkoujok2013application} have used GA and Takagi-Sugeno fuzzy modeling method to detect and isolate sensor faults in nonlinear complex systems.

GAs have been used in modeling reliability of systems. Tian et al. \cite{tian2005line} have used GA to optimize parameters for their ANN in their method for online evaluation of software reliability. In another paper, they have discussed the applicability of their model for predicting software failure time \cite{tian2005evolutionary}. Aljahdali et al. \cite{aljahdali2009software} have discussed a multi-objective GA to assess software reliability in conjunction with other parametric models. 

Meta-heuristic optimization techniques that fall under EC, have been used for system reliability optimization and reliability analysis of various systems. ACO is a comparatively new probabilistic technique that solves combinatorial optimization search problems by selecting good paths through graphs \cite{dorigo1999ant}. Liang et al. have applied for optimal solutions of RAP in series parallel systems \cite{liang2004ant}. Zhao et al. \cite{zhao2007reliability} have developed a multi-objective Ant Colony System (ACS) meta-heuristic for the same problem of redundancy allocation. PSO is meta-heuristic used in reliability analysis as well as optimization of systems in general, and electrical power systems in particular. Robinson et al. \cite{robinson2005reliability} have used PSO to identify critical elements in an electrical grid system. Their method is applicable in reliability analysis of bulk supply systems. Mitra et al.  have used PSO in calculating an optimal load reconfiguration strategy for the power system in an electric ship \cite{mitra2009real}. Bashir et al. have used PSO in the calculation of weights for their adaptive ANN that predicts hourly electric load demand in a grid. Khan et al. \cite{khan2014computational} have used PSO in optimizing their autopilot system for aerospace CPS to improve resilience against faults.  

TS is a metaheuristic optimization technique that attempts to iterate through local optima efficiently with the aim of finding a better optimum in the process. It employs the concept of adaptive memory programming \cite{glover1997tabu} and is suited for large scale problems in reliability analysis where exact solutions are not viable. TS offers an efficient solution for the general optimization of reliability in RAPs. \cite{kulturel2003efficiently}. Caserta et al. \cite{caserta2009tabu} have used TS for software reliability optimization. Other noteworthy uses of TS in CPS related areas can be found in \cite{ramirez2006new} and \cite{pierre1997tabu}. CS is a relatively recent \cite{yang2009cuckoo} optimization algorithm inspired by the parasitic breeding among cuckoos. It is gaining significance, especially for solving redundancy allocation and reliability optimization problems \cite{valian2013improved}. Teske et al. have used CS in locating faults in parallel and distributed systems \cite{teske2015efficient}. Applications of EC in improving system dependability or in modeling dependability are summarized in Table \ref{table:GA}. This table reveals that GAs have been used to solve a variety of problems in optimization and for modeling of CPS dependability. Notable applications in Table \ref{table:GA} include parameter estimation for dependability optimization, redundancy allocation problems, electrical grid reliability optimization and fault prediction.

SA is an iterative search algorithm that was inspired by the physics of annealing of metals \cite{kirkpatrick1983optimization}. It is a probabilistic inference technique \cite{myllymaki1999massively} used to approximate the global optimum of a given function. This technique is particularly suited to find a solution from a large search space. It is efficient in the sense that it incorporates random jumps to potential new solutions. Attiya et al. \cite{attiya2006task} have discussed the problem of task allocation in a heterogeneous distributed system to maximize system reliability using simulated annealing. Similar work by Ravi et al. have discussed the same problem using non-equilibrium simulated annealing \cite{ravi1997nonequilibrium}. Jeon et al. have used a SA based algorithm to optimize power distribution systems \cite{jeon2002efficient}. Fushuan et al. have applied the same technique for fault section estimation in power systems \cite{fushuan1994fault}. Pai et al. \cite{pai2006software} have used SA to calculated parameters for their support vector machine for forecasting software reliability.

\subsection{Artificial Neural Networks}

Based on their biological counterparts, Artificial Neural Networks (ANN) are massively parallel distributed systems for processing information. ANNs can learn from examples. They update previous estimates in light of newly available evidence \cite{bonissone1997soft}. ANNs are comprised of a large number of simple interconnected units that work in parallel to perform a global task. These units can learn and update network parameters in response to an evolving input environment \cite{chaturvedi2008soft}.

ANNs have been used in the analysis and optimization of reliability. They have been applied for parameter estimation for other algorithms. Their learning and prediction capability make them an indispensable tool in robust control and reliability optimization of CPS. 

\begin{table*}[!ht]
\centering
\scriptsize
\caption{Applications of ANN for Improving System Performance and System Dependability Modeling of CPS}
\begin{tabular}{ |m{2.8cm} | m{1.8cm} |m{6cm} |m{2.6cm} |}
\hline
\textbf{Paper Reference}
& \textbf{Soft Computing Technique}
& \textbf{Description}
& \textbf{Dependability Attribute}
\\\hline
  \begin{tabular}[c]{@{}l@{}}Kang et al. 2016\cite{kang2016intrusion}\end{tabular}
 &\begin{tabular}[c]{@{}l@{}}ANN\end{tabular}
&\begin{tabular}[c]{@{}l@{}}Intrusion detection system (IDS) for in-vehicular networks\\ (e.g., CAN)\end{tabular}
&\begin{tabular}[c]{@{}l@{}}Security, Confidentiality\end{tabular}
\\\hline
\begin{tabular}[c]{@{}l@{}}Altiparmak  et al. \\2009 \cite{altiparmak2009general}\end{tabular}
&\begin{tabular}[c]{@{}l@{}}ANN\end{tabular}
&\begin{tabular}[c]{@{}l@{}}Estimation of reliability of telecom network with\\identical link reliability using encoding into ANN\end{tabular}
&\begin{tabular}[c]{@{}l@{}}Reliability, Availability\end{tabular}
\\\hline
\begin{tabular}[c]{@{}l@{}}Cai et al. 2001 \cite{cai2001neural}\end{tabular}
&\begin{tabular}[c]{@{}l@{}}ANN\end{tabular}
&\begin{tabular}[c]{@{}l@{}}Software reliability modeling\end{tabular}
&\begin{tabular}[c]{@{}l@{}}Reliability, Availability\end{tabular}
\\\hline
\begin{tabular}[c]{@{}l@{}}Su et al. 2005 \cite{su2005artificial}\end{tabular}
&\begin{tabular}[c]{@{}l@{}}ANN\end{tabular}
&\begin{tabular}[c]{@{}l@{}}Software reliability assessment and modeling\end{tabular}
&\begin{tabular}[c]{@{}l@{}}Reliability, Availability\end{tabular}
\\\hline
\begin{tabular}[c]{@{}l@{}}Hu et al. 2006 \cite{hu2006early}\end{tabular}
&\begin{tabular}[c]{@{}l@{}}ANN\end{tabular}
&\begin{tabular}[c]{@{}l@{}}Early software reliability prediction\end{tabular}
&\begin{tabular}[c]{@{}l@{}}Reliability, Availability\end{tabular}
\\\hline
\begin{tabular}[c]{@{}l@{}}Hu et al. 2007 \cite{hu2007robust}\end{tabular}
&\begin{tabular}[c]{@{}l@{}}ANN\end{tabular}
&\begin{tabular}[c]{@{}l@{}}Software fault detection, and prediction of correction time\end{tabular}
&\begin{tabular}[c]{@{}l@{}}Reliability, Availability,\\Maintainability \end{tabular}
\\\hline
\begin{tabular}[c]{@{}l@{}}Srivaree et al. 2002 \cite{srivaree2002estimation}\end{tabular}
&\begin{tabular}[c]{@{}l@{}}ANN\end{tabular}
&\begin{tabular}[c]{@{}l@{}}Estimation of all-terminal network reliability\end{tabular}
&\begin{tabular}[c]{@{}l@{}}Reliability, Availability\end{tabular}
\\\hline
\begin{tabular}[c]{@{}l@{}}Bhowmik  et al. 2009 \cite{bhowmik2009novel}\end{tabular}
&\begin{tabular}[c]{@{}l@{}}ANN\end{tabular}
&\begin{tabular}[c]{@{}l@{}}Transmission line fault diagnosis and classification\end{tabular}
&\begin{tabular}[c]{@{}l@{}}Reliability, Availability\end{tabular}
\\\hline
\begin{tabular}[c]{@{}l@{}}Mora et al. 2006 \cite{mora2006fault}\end{tabular}
&\begin{tabular}[c]{@{}l@{}}ANN, FL\end{tabular}
&\begin{tabular}[c]{@{}l@{}}Fault localization in power distribution systems  \end{tabular}
&\begin{tabular}[c]{@{}l@{}}Reliability, Availability\end{tabular}
\\\hline
\begin{tabular}[c]{@{}l@{}}Zhang et al. 2010 \cite{zhang2010artificial}\end{tabular}
&\begin{tabular}[c]{@{}l@{}}ANN\end{tabular}
&\begin{tabular}[c]{@{}l@{}}Load forecasting in smart grids\end{tabular}
&\begin{tabular}[c]{@{}l@{}}Reliability\end{tabular}
\\\hline
\begin{tabular}[c]{@{}l@{}}Gao et al. 2010 \cite{gao2010scada}\end{tabular}
&\begin{tabular}[c]{@{}l@{}}ANN\end{tabular}
&\begin{tabular}[c]{@{}l@{}}SCADA Intrusion Detection and Response Injunction\end{tabular}
&\begin{tabular}[c]{@{}l@{}}Security, Confidentiality\end{tabular}
\\\hline
 \begin{tabular}[c]{@{}l@{}}Linda et al. 2009 \cite{linda2009neural}\end{tabular}
&\begin{tabular}[c]{@{}l@{}}ANNs\end{tabular}
&\begin{tabular}[c]{@{}l@{}}IDS for Critical Infrastructures, SCADA, etc.\end{tabular}
&\begin{tabular}[c]{@{}l@{}}Security, Confidentiality\end{tabular}
\\\hline
\begin{tabular}[c]{@{}l@{}}Moya et al. 2009 \cite{moya2009improving}\end{tabular}
&\begin{tabular}[c]{@{}l@{}}ANNs\end{tabular}
&\begin{tabular}[c]{@{}l@{}}SCADA sensor networks security with Self-Organizing \\Maps (unsupervised ANNs) and reputation systems\end{tabular}
&\begin{tabular}[c]{@{}l@{}}Security, Confidentiality\end{tabular}  
\\\hline
\end{tabular}
\label{tableANN}
\end{table*}

Altiparmak et al. \cite{altiparmak2009general} have used ANNs to model the reliability of communication networks with links that have identical reliability. The node and link can vary in size in their model. Srivaree et al. \cite{srivaree2002estimation} have used ANNs to learn from existing topologies and predict network reliability in an all-terminal network. Bhowmik et al. \cite{bhowmik2009novel}  have used ANNs in conjunction with discrete wavelet transform (DWT) to predict and classify transmission line faults. Zhang et al. \cite{zhang2010artificial}  have used ANNs to forecast load demand in smart grids. Mora et al. \cite{mora2006fault} have used neuro fuzzy classifiers for locating faults in smart grids. ANNs have been used to analyze and forecast the reliability of software. Cai et al. \cite{cai2001neural} have discussed 
the effectiveness of neural networks for handling dynamic software reliability data. Other noticeable works in this domain include Yu-Shen Su et al. \cite{su2005artificial} , Hu et al. \cite{hu2006early,hu2007robust}, Y Singh et al. \cite{singh2010application}.

ANNs have been used in combination with optimization techniques like GA to estimate initial values for optimization. ChangYoon Lee et al. \cite{changyoon2002reliability} have proposed a hybrid GA/ANN with FL logic controller for RAP.  The learning capability of ANNs makes them particularly suited for IDS. They also have found multiple applications in computer networks, SCADA systems, smart grids and other CPS or CPS related systems. Gao et al. \cite{gao2010scada} discussed an IDS for smart utilities that uses an ANN with three stage back propagation. Linda et al. \cite{linda2009neural} have used a supervised ANN based IDS for power grid applications. They use Levenberg-Marquardt algorithm and back error propagation to train their network. Kang et al. \cite{kang2016intrusion} have used Deep Neural Network (DNN) structure for intrusion detection in order to improve the security of in-vehicular networks (e.g., CAN: Controller Area Network). Their technique uses high-dimensional CAN packet data to train their deep belief network which can differentiate attacked packets from normal ones based on their statistics. Moya et al. \cite{moya2009improving} have used Self Organizing Maps (SOM) for improving the security of sensor data in SCADA systems. Applications of ANN for dependability analysis or optimization in CPS are summarized in Table \ref{tableANN}. A glance at Table \ref{tableANN} indicates that ANN have been used mainly for early fault prediction, fault localization, and intrusion detection.

\subsection{Probabilistic Reasoning}

Probabilistic Reasoning (PR), also referred to as probabilistic inference and probabilistic logic in the literature — deals with uncertainty and belief propagation \cite{zadeh1994soft}. PR is a formal mechanism to apply probability theory and subsidiary techniques for decision making under uncertainty. It allows BN cluster analysis and analysis of stochastic systems \cite{zadeh1998some}. The term probabilistic in PR hints at the reasoning mechanisms and probabilistic representations grounded in probability theory 
\cite{de2003probabilistic} and the Dempster-Shafer theory of evidence \cite{he2006probabilistic}. PR subsumes Belief Networks, Chaos Theory and parts of machine learning theory \cite{shiu2004case}. Graphical methods like Markov Logic Networks (MLN) (also known as Markov Random Fields, MRF) also fall under this category \cite{kindermann1980markov,richardson2006markov}. The emerging paradigm of probabilistic programming and probabilistic programming languages provide a formal framework to apply probabilistic inference to uncertainty related problems \cite{gordon2014probabilistic}.

 \begin{table*}
\centering
\scriptsize
\caption{Applications of PR for Improving System Performance and System Dependability Modeling of CPS}
\begin{tabular}{ |m{2.8cm} | m{1.8cm} |m{6cm} |m{2.9cm} |}
\hline
\textbf{Paper Reference}
& \textbf{Soft Computing Technique}
& \textbf{Description}
& \textbf{Dependability Attribute}
\\\hline
 \begin{tabular}[c]{@{}l@{}}Weber et al. 2006 \cite{weber2006complex}\end{tabular}
&\begin{tabular}[c]{@{}l@{}}BN\end{tabular}
&\begin{tabular}[c]{@{}l@{}}Dynamic modeling of complex manufacturing processes\\ using Dynamic Object-Oriented Bayesian Networks\\ (DOOBNs). DOOBN (with FTA) used for dependability \\modeling.\end{tabular}
&\begin{tabular}[c]{@{}l@{}}Reliability\end{tabular}
\\\hline
\begin{tabular}[c]{@{}l@{}}Weidl et al. 2005 \cite{weidl2005applications}\end{tabular}
&\begin{tabular}[c]{@{}l@{}}BN\end{tabular}
&\begin{tabular}[c]{@{}l@{}}Object-Oriented Bayesian Networks (OOBNs) for isolation \\of faults in complex industrial systems and for decision \\support\end{tabular}
&\begin{tabular}[c]{@{}l@{}}Reliability, Availability,\\Maintainability\end{tabular}
\\\hline
\begin{tabular}[c]{@{}l@{}}McNaught  et al. 2009\cite{mcnaught2009using}\end{tabular}
&\begin{tabular}[c]{@{}l@{}}BN\end{tabular}
&\begin{tabular}[c]{@{}l@{}}Prognostic Modeling and Maintenance Decision Making.\\ Dynamic BNs for dependability modeling\end{tabular}
&\begin{tabular}[c]{@{}l@{}}Reliability, Maintainability\end{tabular}
\\\hline
\begin{tabular}[c]{@{}l@{}}Huang et al. 2006\cite{huang2006bayesian}\end{tabular}
&\begin{tabular}[c]{@{}l@{}}BN, FL, GA\end{tabular}
&\begin{tabular}[c]{@{}l@{}}Bayesian reliability analysis with parameters found using\\ FL and GA. Estimation of pdfs of reliability\\ using FL and Bayesian analysis\end{tabular}
&\begin{tabular}[c]{@{}l@{}}Reliability, Availability\end{tabular}
\\\hline
\begin{tabular}[c]{@{}l@{}}Wang et al. 2009 \cite{wang2009bayesian}\end{tabular}
&\begin{tabular}[c]{@{}l@{}}BN\end{tabular}
&\begin{tabular}[c]{@{}l@{}}Reliability Analysis from incomplete and insufficient data \\sets. BNs for dependability modeling\end{tabular}
&\begin{tabular}[c]{@{}l@{}}Reliability, Availability\end{tabular}
\\\hline
\begin{tabular}[c]{@{}l@{}}Liu et al. 2009 \cite{liu2009scalability}\end{tabular}
&\begin{tabular}[c]{@{}l@{}}BN\end{tabular}
&\begin{tabular}[c]{@{}l@{}}Quantification of scalability of network  resilience upon \\failures. BNs used for dependability modeling.\end{tabular}
&\begin{tabular}[c]{@{}l@{}}Reliability, Availability,\\Maintainability, Survivability
\end{tabular}
\\\hline
\begin{tabular}[c]{@{}l@{}}Queiroz et al. 2013 \cite{queiroz2013probabilistic}\end{tabular}
&\begin{tabular}[c]{@{}l@{}}BN\end{tabular}
&\begin{tabular}[c]{@{}l@{}}Modeling and quantification of overall resilience of \\networked systems. MLN and MRF used for dependability\\ modeling\end{tabular}
&\begin{tabular}[c]{@{}l@{}}Reliability, Availability,\\Maintainability, Survivability
\end{tabular}
\\\hline  
\end{tabular}
\label{tablePR}
\end{table*}

Recent literature reveals a growing interest in reliability modeling using BNs, particularly to complex systems \cite{langseth2008bayesian}. BNs estimate the distribution probabilities of a set of variables based on observation of some variables and prior knowledge of others. BNs allow us to merge knowledge of diverse nature into a single data \cite{weber2012overview}. This is particularly suitable for complex systems. BNs establish cause effect relationships and model their interactions.  Weber et al.\cite{weber2012overview} have reviewed applications of BNs in dependability and risk analysis, and maintenance. They report an 800\% increase in interest in the use of BNs for dependability analysis. 

BNs can be used to represent local dependencies as well as for predictive and diagnostic reasoning. BNs are superior to classical methods like FT analysis of complex systems \cite{boudali2005new}. Bobbio et al. \cite{bobbio2001improving} presented an algorithm for mapping FTs into BNs. Montani et al. \cite{montani2008radyban} have developed a software for this purpose. A formal analysis of this conversion for dynamic fault trees was discussed in \cite{boudali2005discrete}. In most engineering problems, known statistics about the reliability of a component or systems are insufficient for predicting their random behavior. Further subjective human analysis needs to be considered. Wang et al. \cite{wang2009bayesian} have used BNs for reliability modeling and prediction with subjective data sets with insufficient or incomplete information.

Weber et al. \cite{weber2006complex} have introduced Dynamic Object Oriented Bayesian Networks (DOOBNs) as an alternative technique to conventional reliability analysis tools like MC and FTA for modeling the reliability of complex industrial systems. Object oriented version of BN allows for elegant, smaller representation of otherwise complex BNs. BNs are suitable for the modeling of the propagation of failures in a complex system \cite{weber2006complex}  because of the way they capture cause and effect relationships. Weidl et al. \cite{weidl2005applications} have used Object Oriented Bayesian Networks (OOBNs) for isolation of faults in complex industrial systems and for decision support. They have used BNs to handle uncertainty in measured sensor data. McNaught et al. \cite{mcnaught2009using} have discussed dynamic BNs in the prognostic modeling of a component's state.  Guanglei Liu et al. \cite{liu2009scalability} have used Bayesian Bayesian networks to model network failure. BNs show dependencies among different link failures explicitly.

An MLN, or Markov Random Field (MRF), is a probabilistic logic that applies the concepts of a Markov Network (MN) to first order logic. It is similar to a BN in the representation of dependencies. However, BNs are directed and acyclic, whereas MNs are undirected and may even be cyclic. An MN can, therefore, represent cyclic dependencies, something not possible with a Bayesian network. On the flip side, it cannot represent dependencies such as induced dependencies that are possible with Bayesian networks. Queiroz et al. \cite{queiroz2013probabilistic} have used MN to model and quantify overall resilience of networked systems on the basis of inter-dependencies of services and their adaptation. Applications of PR in terms of system dependability modeling and optimization are summarized in Table \ref{tablePR}. The table shows that BNs are by far the most used PR technique. PR has also been used for modeling of dependability and for prediction of faults in a variety of systems.

\subsection{Rough Set Theory}

Introduced in 1982, Rough Set Theory (RST) is a relatively new approach for data analysis and inference in the presence of vagueness and uncertainty \cite{pawlak2012rough}. RS theory is another method for analyzing uncertain systems and is gaining interest as a tool for knowledge discovery \cite{frias2005modeling}, data mining, classification and image processing. It provides a systematic framework for dealing with vagueness caused by indiscernibility when complete information about a system is not available \cite{yao1998comparative}. Rough sets are a non-invasive method of knowledge discovery \cite{duntsch2000rough}. They need minimal model assumptions and can usually determine all parameters from within the observed data. This alleviates the need for other information like grade of membership and values for possibility required by fuzzy set theory \cite{rissino2009rough}. RS theory can help in the construction of models that represent the underlying domain theory from a set of data alone \cite{shiu2004case}. Rough and Fuzzy set theories are different approaches to handle vagueness that attempt to remedy the difficulties with classical set theory \cite{pawlak1982rough}. They were an attempt at the generalization of classical set theory so that vagueness and uncertainty could be modeled \cite{yao1998comparative}.

RS based analysis provides a self-contained framework that can potentially obviate the need for external information such as a priori distributions in statistical analysis, model assumptions, or membership grade in fuzzy set theory. The core of  RST is to weigh attributes by importance and reduce their total number \cite{li2009improved}. RST has been applied to analyze the dependability of a wide range of systems. It has been used generously in reliability analysis of electrical power systems and mechanical systems, with some recent applications in software systems. 
\begin{table*}
\centering
\scriptsize
\caption{Applications of Rough Set Theory for Improving System Performance and System Dependability Modeling of CPS}
\begin{tabular}{ |m{2.8cm} | m{1.8cm} |m{5.8cm} |m{2.6cm} |}
\hline
\textbf{Paper Reference}
& \textbf{Soft Computing Technique}
& \textbf{Description}
& \textbf{Dependability Attribute}
\\\hline
\begin{tabular}[c]{@{}l@{}}Peng et al. 2004 \cite{peng2004rough} \end{tabular}
&\begin{tabular}[c]{@{}l@{}}RST\end{tabular}
&\begin{tabular}[c]{@{}l@{}}Data mining for fault diagnosis in (electric power)\\distribution feeders. RS used for dependability modeling. \end{tabular}
&\begin{tabular}[c]{@{}l@{}} Reliability, Availability,\\Maintainability\end{tabular}
\\\hline 
\begin{tabular}[c]{@{}l@{}}H Su et al. 2005\cite{su2005substation}\end{tabular}
&\begin{tabular}[c]{@{}l@{}}RST, ANN\end{tabular}
&\begin{tabular}[c]{@{}l@{}}Substation fault diagnosis based on RST and ANN model \\(in electric power systems)\end{tabular}
&\begin{tabular}[c]{@{}l@{}}Reliability, Availability,\\Maintainability
\end{tabular}
\\\hline
\begin{tabular}[c]{@{}l@{}} Chen et al. 1998 \cite{chen1998software}\end{tabular}
&\begin{tabular}[c]{@{}l@{}}RST\end{tabular}
&\begin{tabular}[c]{@{}l@{}}Software safety evaluation (for  safety-critical systems)\\\end{tabular}
&\begin{tabular}[c]{@{}l@{}}Reliability, Safety\end{tabular}
\\\hline
\begin{tabular}[c]{@{}l@{}}Li, Bo, and Yang \\Cao 2009\cite{li2009improved}\end{tabular}
&\begin{tabular}[c]{@{}l@{}}RS\end{tabular}
&\begin{tabular}[c]{@{}l@{}}A comprehensive  model for software dependability\\evaluation using RST. RST based modeling\\of dependability attributes\end{tabular}
&\begin{tabular}[c]{@{}l@{}}Availability, Reliability,\\Safety, Confidentiality,\\Integrity, Maintainability \end{tabular}
\\\hline
\begin{tabular}[c]{@{}l@{}}Yuan et al. 2007\cite{yuan2007rough}\end{tabular}
&\begin{tabular}[c]{@{}l@{}}RS\end{tabular}
&\begin{tabular}[c]{@{}l@{}}Multi-state system reliability estimation using RST and\\Petri Nets. SPN  and RST for dependability modeling\end{tabular}
&\begin{tabular}[c]{@{}l@{}}Reliability\end{tabular}
\\\hline

\begin{tabular}[c]{@{}l@{}}Wang et al. 2004 \cite{wang2004rough}\end{tabular}
&\begin{tabular}[c]{@{}l@{}}RS\end{tabular}
&\begin{tabular}[c]{@{}l@{}}A rough set-based prototype system that aims at ranking\\ the possible faults for fault diagnosis\end{tabular}
&\begin{tabular}[c]{@{}l@{}}Reliability, Maintainability\end{tabular}
\\\hline

\begin{tabular}[c]{@{}l@{}}Joslyn et al. 2003\cite{joslyn2003multi}\end{tabular}
&\begin{tabular}[c]{@{}l@{}}RS\end{tabular}
&\begin{tabular}[c]{@{}l@{}}Construction of random intervals for reliability analysis\end{tabular}
&\begin{tabular}[c]{@{}l@{}}Reliability\end{tabular}
\\\hline
\begin{tabular}[c]{@{}l@{}}Song et al. 2014\cite{song2014rough}\end{tabular}
&\begin{tabular}[c]{@{}l@{}}RS\end{tabular}
&\begin{tabular}[c]{@{}l@{}}Reliability Modeling using FMEA and RST in  uncertain\\ environments\end{tabular}
&\begin{tabular}[c]{@{}l@{}}Reliability, Availability\end{tabular} 
\\\hline

\end{tabular}

\label{tableRS}
\end{table*}
Li et al. \cite{li2009improved} have presented a comprehensive evaluation model for software dependability using RS. The earlier fuzzy model required objective weight calculation from statistical data on software dependability. In a newer method proposed by Li et al., an approach is proposed that uses a combination weight that takes expert's subjective knowledge as well in addition to objective data. In particular, objective weight is calculated from statistical data using RST.

The ability of RST to analyze non-quantitative information was exploited in a paper by Chen-Jimenez et al. \cite{chen1998software}. They used RST to quantify the safety of software for safety-critical systems. Verbal subjective judgments from users were used as the input to the system. Wang et al. \cite{wang2004rough} have presented an RST based system to rank faults in order to optimize maintenance work by ordering faults. RST is used as a knowledge extraction tool to learn from and analyze past fault diagnosis records, diagnosis from experts and to extract a set of minimal diagnostic rules. The system then uses RS sets to rank or order these faults. Joslyn \cite{joslyn2003multi} et al. have discussed RS analysis to calculate random intervals from simple multi-intervals. Such intervals are required for some reliability analysis techniques \cite{qiu2008probabilistic,cui2013multi}. The aim of such analysis is to obtain the system failure probability interval from available statistical parameter intervals of the underlying variables \cite{qiu2008probabilistic}. Random intervals offer the advantage of representing randomness via probability theory while imprecision and non-specificity via intervals at the same time. This can complement probabilistic analysis with other techniques such as FL, plausibility and belief measure \cite{joslyn2003multi}. RST allows researchers to construct complex random interval representations and elicit simple multi-interval information. RST has found extensive use in data mining to infer patterns and rules for making predictions. Peng et al. \cite{peng2004rough} have used RS to predict distribution feeder faults and for fault localization by modeling causal relationships between faulty equipment and evidence provided by observations. Applications of RST in CPS dependability analysis and optimization are summarized in Table \ref{tableRS}. This table shows that RST has been used mainly for modeling of dependability and also as a data analysis technique for fault prediction.

\section{Insights: Recommendations and Pitfalls}

An extensive study of literature, as summarized in Tables \ref{table:FL}---\ref{tableRS}, reveals that among the attributes of dependability, \textit{reliability} and \textit{availability} have found the most applications. This is followed by \textit{maintainability} (i.e., fault tolerance and repairability) and \textit{confidentiality} (security). It was also noted that soft computing techniques have been used mostly for optimization of performance or reliability of systems. Soft computing has been used in aiding reliability and dependability analysis of systems as well. Soft computing cannot be a substitute for other rigorous methods of reliability analysis. In most instances, soft computing has been used in classification or to dig out extra information about the dependability of a system or to approximate reliability measures. 

Each soft computing technique has particular computational parameters that make it suitable for a particular problem. Fuzzy set theory has been used primarily as a substitute for the classical probability to represent the availability or reliability of a system or component in terms of possibilities instead of probabilities. This has been applied in reliability analysis of traditional or mechanical systems \cite{karwowski1986potential,bing2000practical,kai1991fuzzy}. The application of FL in fault tree analysis garnered interest in the 90s. Again, it has mostly been applied to conventional systems \cite{yuhua2005estimation,shu2006using}. The inclusion of fuzzy possibilities in reliability analysis adds some computational overhead, compared to more classical methods involving probability measures. FL has been used in reliability optimization and resource allocation problems to some extent \cite{ravi2000fuzzy,mahapatra2006fuzzy,mahapatra2014reliability}. A noticeable use of fuzzy logic in modern systems has been in classification and fuzzy inference \cite{huiling2008software,ebrahimipour2013emotional,knezevic2001reliability,tyagi2014adaptive,toosi2007new}. Fuzzy logic has also been used in the estimation of parameters for other soft computing or ML techniques \cite{lin2006self,cho2002incorporating,huang2006bayesian}. FIS allow easy encoding of a priori knowledge. However, they rely on expert judgment for the design of fuzzy rules and membership functions. Generally, fuzzy inference is used in conjunction with an ML technique like ANNs and GAs to learn parameters for the inference engine.

EC offers a host of optimization algorithms suited to reliability optimization and RAP. Such problems generally arise in the design phase of systems. EC has also been used in localization of faults \cite{elkoujok2013application,teske2015efficient,tian2005evolutionary,robinson2005reliability}. PSO has been widely used in optimization of smart grids and power system \cite{mitra2009real,robinson2005reliability,bashir2009applying}. SA is used primarily as an optimization technique for redundancy allocation \cite{attiya2006task,jeon2002efficient}. There are very few instances of direct use of EC in a holistic reliability model or analysis for a particular system. 

Neural networks are, by far, the most diverse soft computing technique. They find applications in reliability analysis, optimization, and prediction. ANNs have also been widely used in conjunction with other soft computing methods. ANNs have been used in reliability modeling \cite{altiparmak2009general} and even to predict the reliability of systems \cite{bhowmik2009novel}. Their ability for prediction has been exploited in systems as well \cite{zhang2010artificial}. Its learning ability has been put to use to localize sources of faults from available data \cite{bhowmik2009novel} and in IDS \cite{gao2010scada,kang2016intrusion,moya2009improving}. ANN are good at pattern recognition but not at explaining how they reach their decisions. FL uses a relatively simple construction. It is easier to comprehend their decision process but FL alone cannot draw the rules for making decisions. ANNs have  been used in conjunction with other soft computing techniques like FL in reliability analysis and optimization \cite{mora2006fault,tyagi2014adaptive}. ANNs can be used for evaluation of reliability. However, a large training data set is required for accurate analysis and this increases computational cost.

The primary probabilistic reasoning techniques discussed here is BN. BNs have been used extensively in reliability analysis and modeling \cite{wang2009bayesian,weber2006complex} as well as a predictive tool for component reliability \cite{liu2009scalability}. Probabilistic programming is an emerging inference tool that facilitates the application of statistical analysis and offers all kinds of potential applications including reliability analysis and optimization \cite{gordon2014probabilistic}.

RS have found uses in data classification, mining, and inference with the primary emphasis on inference from a limited set of untagged data. Their application as a precise tool for reliability analysis in the classical sense is limited. However, they can provide a qualitative method to categorize faults \cite{wang2011secured,peng2004rough}. They can be used in the calculation of random intervals \cite{qiu2008probabilistic,cui2013multi} in order to complement other reliability analysis techniques.

\section{Open Issues and Future Works}

While there is plenty of literature on dependability analysis of electrical, mechanical and even networked systems that make up a CPS, we find a general lack of literature specific to dependability analysis of CPS or synthesis of reliability analysis for CPS in terms of its components. The need for such work will increase as efforts to standardize the architecture for CPS gathers momentum \cite{lee2015cyber}. 

\subsection{Lack of a Unified Modeling or Analysis Framework}
The design of CPS is challenging in terms of physical systems and hardware, and even in a programming language to implement the desired level of computational behavior. A unified framework is required to model the component of CPS, which allows easy interfacing and consistency. Such a framework should allow modeling of the physical environment, support heterogeneity, and scalability, and integrate well with existing simulation and verification tools \cite{wan2010specification}. This will cause effective modeling of asynchronous dynamics by integrating event and time-based computation. 

\subsection{Design Methodologies}
CPS are being deployed on a wide scale in diverse kinds of applications. Many systems including smart homes and power systems are being operated in new ways that were never intended for them \cite{khaitan2015design}. Novel design methodologies are required for their seamless integration with new systems while avoiding disruption in new systems, and also to ensure dependable operation while providing new extensions of capabilities.    
\subsection{Security}
One of the major obstacles that CPS must overcome is ensuring security while maximizing mutual coordination among cyber and physical components. Reliability and security are very crucial in mission-critical CPS like healthcare, smart power grids and networking systems. Future CPS must operate with enhanced security and reliability. There is a crucial need to develop such intelligent architectures that can ensure real-time security-state monitoring and remediation. Security performance metrics must be developed and standardized to evaluate the security of the systems. Security in CPS is a real concern since the feedback loop signals and control commands are often transported over the public networks and use open standards \cite{cardenas2008research} in order to minimize costs. Intrusion Detection Systems (IDS) is an active area of research in CPS dependability. Researchers plan to improve the CPS survivability by modeling and predicting attacks using game theory \cite{mitchell2016modeling}. 
\subsection{Network Induced Constraints}
Reliability in large scale (complex) network control systems (NCS) is often difficult to model because of unpredictable random delays in the underlying communication links. Current control, communications, and software theory have not matured enough to solve problems caused by the heterogeneity in CPS. CPS can contain control loops separated by geographical scale distances. The impact of the network on closed-loop system performance \cite{walsh2002stability}, stability \cite{li1997criteria} and ultimately reliability is another area that remains to be looked at. The significance of combining control specifications and communication constraints has not  been addressed \cite{walsh2002stability}. NCS must cope with network induced constraints. In the literature, five types of network induced constraints have been identified. These include packet losses and disorder, time-varying transmission intervals, competition among different nodes for accessing the same network, time delays, and data quantization delays  \cite{zhang2013network}. Delays in networked control systems cannot be modeled using conventional delay systems since data is transmitted in packets and scheduled through a system which is generally designed to package large amounts including the sequence of control commands. Comprehensive studies combining these constraints are not available \cite{zhang2013network}. The role of the network in closed-loop system performance \cite{walsh2002stability}, stability \cite{li1997criteria} and ultimately reliability remains to be explored in depth. Inserting a network in a control loop may cause deteriorated system performance or even instability \cite{zhang2013network}. A unified theory of heterogeneous control and communication systems would help in this regard \cite{wang2008cyber}. Efforts to this end must also contend with the complicated security challenges posed by CPS.
\subsection{Soft Computing in the Control Loop}
Soft computing is being used in improving the stability and fault tolerance of control systems. Control reconfiguration is an active approach for fault tolerant control of dynamic systems \cite{zhang2008bibliographical}. Soft computing techniques like FL and ANNs have been used in control of such adaptive systems while GAs have been used to design fault-tolerant systems. Fault-tolerant control impacts the reliability modeling and assessment of systems \cite{blanke1997fault}. A discussion on soft computing directly in the control loop is another avenue to explore CPS dependability. 
\subsection{Distributed Collaborative Control}
Distributed collaborative control in an unreliable wireless network \cite{chen2010distributed} is yet another area where reliability analysis could be explored. The merger of reliability analysis and soft computing with modern research on distributed control systems would aid in designing more dependable CPS.   
\subsection{Probabilistic Computing and CPS}
The new paradigm of probabilistic computing offers a host of tools that will eventually facilitate reliability analysis. While the proponents of probabilistic programming have pointed out its use for this purpose \cite{gordon2014probabilistic}, the literature on the subject is almost non-existent. 
\subsection{Standardization Requirements}
CPS applications depend on multiple advanced technologies from multiple industries. The scope of standardization required for CPS is much larger than any traditional standards development \cite{tyagi2016cyber}. These standardization efforts must inevitably address the stringent Quality of Service (QoS) and dependability requirements for CPS.

\section{Conclusion}
 In this paper, we provide a comprehensive in-depth review of the applications of soft computing for dependability analysis and dependability improvement of CPS and similar systems. The diversity and extent of areas that must be covered for such a survey compounded by the classification and labeling of soft computing techniques applicable to the subject makes it look like a daunting task. Nevertheless, we summarize applicable domains and scenarios where one or more soft computing technique has been used in reliability analysis or optimization. This study reveals a significant lack of literature available on comprehensive reliability analysis or optimization of CPS. Given the tremendous opportunities CPS will offer in the foreseeable future and given the interest in the applications of soft computing in recent years, it is only natural to conclude that interest in the subject explored in this survey will only grow with time.




\end{document}